\documentclass[12pt]{elsarticle}

\usepackage{amsmath}
\usepackage{graphicx} % Allows for eps images
\usepackage{siunitx}
\usepackage{natbib}
\usepackage{float}
\usepackage[a4paper, margin=1in]{geometry}

\makeatletter
\def\ps@pprintTitle{%
  \let\@oddhead\@empty
  \let\@evenhead\@empty
  \def\@oddfoot{\reset@font\hfil\thepage\hfil}
  \let\@evenfoot\@oddfoot
}
\makeatother

\usepackage{amssymb}
\begin{document}

\begin{frontmatter}



\title{Thermal resistance from non-equilibrium phonons at Si-Ge interface}


\address[1]{Department of Mechanical Engineering and Materials Science, University of Pittsburgh, Pittsburgh, PA 15261, USA}
\address[2]{Department of Physics and Astronomy, University of Pittsburgh, Pittsburgh, PA 15261, USA}
\fntext[ref1]{These authors contributed equally to this work}
\fntext[ref2]{current address: Materials Science and Technology Division, Oak Ridge National Laboratory, Oak Ridge, TN 37831, USA}

\author[1]{Xun Li\fnref{ref1,ref2}}
\author[1]{Jinchen Han\fnref{ref1}}
\author[1,2]{Sangyeop Lee\corref{cor1}}
\cortext[cor1]{sylee@pitt.edu}

\begin{abstract}
As nanostructured devices become prevalent, interfaces often play an important role in thermal transport phenomena. However, interfacial thermal transport remains poorly understood due to complex physics across a wide range of length scales from atomistic to microscale. Past studies on interfacial thermal resistance have focused on interface-phonon scattering at the atomistic scale but overlooked the complex interplay of phonon-interface and phonon-phonon scattering at microscale. Here, we use the Peierls-Boltzmann transport equation to show that the resistance from the phonon-phonon scattering of non-equilibrium phonons near a Si-Ge interface is much larger than that directly caused by the interface scattering. We report that non-equilibrium in phonon distribution leads to significant entropy generation and thermal resistance upon three-phonon scattering by the Boltzmann’s H-theorem. The physical origin of non-equilibrium phonons in Ge is explained with the mismatch of phonon dispersion, density-of-states, and group velocity, which serve as general guidance for estimating the non-equilibrium effect on interfacial thermal resistance. Our study bridges a gap between atomistic scale and less studied microscale phenomena, providing comprehensive understanding of overall interfacial thermal transport and the significant role of phonon-phonon scattering.

\end{abstract}



%\begin{keyword}



%\end{keyword}

\end{frontmatter}


\newpage
\section{Introduction}
\label{Introduction}





Thermal transport across interfaces between solid materials has drawn significant interest due to its importance in applications including thermal management and energy conversion processes\cite{RN1,RN2,RN3,RN4}. The fast heat dissipation has become one of the greatest challenges that is crucial for various devices from integrated circuits to massive data centers\cite{RN4,RN5}. As the density of interfaces increases, thermal transport is determined by not only the intrinsic properties of the bulk material but also conditions of thermal interfaces. In these cases, the resistance caused by thermal interfaces can be larger than thermal resistance of the material itself and plays a key role in thermal transport. Due to the complexity around thermal interfaces such as atomic structure mismatching, interaction among heat carriers, etc., a better understanding of interfacial resistance is still the center of recent research efforts\cite{RN3,RN4,RN6}.

In recent years, many progresses have been made on the theory and simulation of interfacial thermal transport, mostly focusing on the interfacial scattering at atomistic scale. The conventional theories, such as acoustic mismatch model (AMM) and diffuse mismatch model (DMM), have drawbacks as they predict interfacial phonon scattering based on bulk properties of two constituent materials and do not consider the effects of local atomic structure and bonding strength on interfacial thermal transport. Recent atomistic simulations, for example, the atomistic Green’s function (AGF) and molecular dynamics (MD) simulation, overcome these drawbacks. The AGF finds spectral phonon transmission function and has been widely used for various types of interfaces\cite{RN7,RN8,RN9}. The AGF was recently extended to the mode-resolved AGF that can predict modal phonon transmission function\cite{RN10,RN11}. In addition, while the previous AGF assumes only harmonic processes for energy transfer, anharmonic AGF was developed to include inelastic interfacial scattering\cite{RN12,RN13,RN14}. The MD simulations naturally include atomic structures at the interface and anharmonicity. Recent MD studies have identified interfacial phonon modes\cite{RN15,RN16,RN17} and obtained transmission function based on phonon modes\cite{RN17,RN18}.

Although those MD and AGF have significantly advanced the understanding on detailed mechanisms of interfacial phonon transport at atomistic scale, they have limited capabilities for phenomena occurring at larger length scale, for example the combined effects of phonon-interface and phonon-phonon scattering within a few micrometers from the interface. The phonon-phonon scattering can be important for interfacial thermal transport in two aspects. First, the distribution of phonons encountering the interface is determined by the balance of phonon advection and phonon scattering including the phonon-phonon scattering. Second, the phonon-phonon scattering relaxes highly non-equilibrium phonons near an interface and should cause thermal resistance during the relaxation process as further discussed below.

The phonon distribution at the interface is necessary to predict modal and total interfacial thermal resistance in the atomistic simulations. The MD simulations usually include thermal reservoirs that emit phonons with equilibrium distribution to leads. The emitted phonons propagate through the leads experiencing phonon-phonon scattering, but the typical length of leads in the MD simulation is much shorter than the phonon mean free paths. In such cases, phonons arriving at the interface are expected to have a distribution function close to the equilibrium one. The AGF is usually combined with the Landauer formalism\cite{RN19} assuming that phonons emitted with the equilibrium Bose-Einstein distribution ($f^{\text{eq}}$) travel through a lead without phonon-phonon scattering. The equilibrium phonon distribution assumed in the MD and Landauer formalism can be reasonable when the flux through the leads is very small compared to that through the interface (or device) and thus carrier distribution in the leads does not deviate much from the equilibrium distribution. A typical example is the transport through a nanowire placed between two large leads. Since the leads have a much larger cross-sectional area than the nanowire, the flux in the lead can be negligibly small and thus the carrier distribution in the lead can be reasonably assumed as the equilibrium one. However, for interfacial thermal transport, the cross-sectional area of the interface is the same as that of leads, and thus the carrier distribution at the lead can largely deviate from the equilibrium one. Assuming two semi-infinite leads sharing an interface, the phonon distribution far from the interface should be the same as that in an infinitely long lead with a homogenous temperature gradient which we call bulk distribution, $f^{\text{bulk}}$. For this reason, previous studies suggested to use $f^{\text{bulk}}$ instead of $f^{\text{eq}}$ for the Landauer formalism\cite{RN20,RN21}. However, such modified Landauer formalism still has a limit; the actual distribution of phonons encountering the interface can be different from $f^{\text{bulk}}$ due to combined effects of the phonon-interface and phonon-phonon scattering. Thermal resistances from the modified Landauer formalism exhibit large disagreement with non-equilibrium molecular dynamics (NEMD) simulations and experimentally measured resistance\cite{RN22}. Also, it was suggested that the EMD and NEMD simulations correspond to the Landauer formalism using the equilibrium and bulk phonon distributions, respectively\cite{RN23}.

We hypothesize that the second effect of the phonon-phonon scattering, the relaxation of non-equilibrium phonons caused by the interface scattering, can be particularly important for the interfacial thermal resistance. In the two semi-infinite leads sharing an interface, the phonon distribution far from the interface is $f^{\text{bulk}}$, moderately deviated from the equilibrium distribution and thus causing non-zero heat flux. However, the phonons near the interface should be highly non-equilibrium due to the reflection and transmission occurring at the interface, which severely distort the distribution function such that the distribution near the interface is much different from $f^{\text{bulk}}$. Recent MD simulation with mode resolution reports that the temperature of each mode significantly differs from others implying highly non-equilibrium phonon distribution near the interface\cite{RN18}. Thus, as phonons diffuse from the interface to the leads or vice versa, phonon distribution should gradually change upon the phonon-phonon scattering in the lead. During such changes of phonon distribution, the phonon-phonon scattering process should generate entropy and thermal resistance by the Boltzmann’s H-theorem. Even when the phonon-phonon scattering is the momentum-conserving type which is usually considered not to directly cause thermal resistance, the scattering process generates considerable amount of thermal resistance when the distribution changes from equilibrium to the bulk distributions\cite{RN24,RN25}. Considering the large difference between bulk and interface phonon distributions and the high degree of non-equilibrium, the resistance due to the distribution change by the phonon-phonon scattering is expected to be significant. However, the simple Landauer theory is not capable of predicting the resistance from the scattering of non-equilibrium phonons near an interface since it assumes ballistic transport of phonons without scattering.

In this paper, we examine the combined effect of phonon-interface scattering at a Si-Ge interface and phonon-phonon scattering in the Si and Ge leads on the overall interfacial thermal resistance. We solve the steady-state Peierls-Boltzmann transport equation (PBE) using a kinetic Monte Carlo (MC) technique for phonon transport across the interface of semi-infinitely long Si and Ge leads. With inputs of phonon dispersion and scattering matrices from first principles and the transmission function from the DMM, the MC solver provides the phonon distribution function in both real and reciprocal spaces, capturing the significant change of distribution near the interface due to the phonon-phonon scattering. In particular, we calculate the local entropy generation from the phonon-phonon scattering and quantitatively analyze the thermal resistance from the scattering of non-equilibrium phonons near the interface.

\newpage
\section{Methods}
\label{Methods}

\begin{figure}[H]
\includegraphics[width=\textwidth]{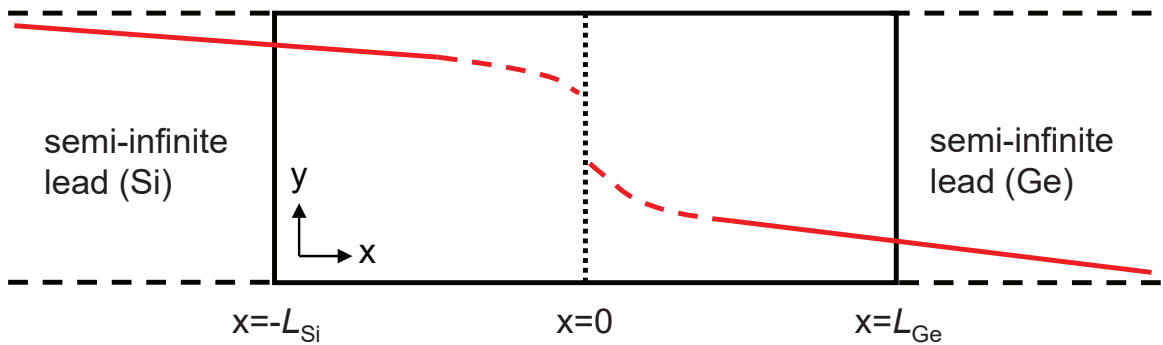}
\caption{Schematic picture of the Si-Ge interface (black dotted line) shared by two semi-infinite leads (black dashed lines) and the finite-sized computational domain (black solid lines). The red line represents the sketch of temperature profile showing the constant temperature gradients far from the interface.}
\label{fig:1}
\end{figure}

We study the thermal transport across an interface by solving the PBE with a relaxation time approximation (RTA) for the three-phonon scattering:
\begin{align} \label{eq:PBE1}
v_{i} \nabla f_{i}(x) = -\frac{f_{i}(x) - f_{i}^{\text{loc}}(x)}{\tau_{i}}
\end{align}
where $f_{i}(x)$ is the distribution of phonon mode $i$ at location x, $f^{\text{loc}}$ is the Bose-Einstein distribution at local equilibrium temperature $T_{\text{loc}}$, $v$ is the phonon group velocity, and $\tau$ is the phonon lifetime. When both phonon-interface and three-phonon scatterings are considered, the thermal interfacial resistance is an extrinsic property of the interface as the resistance value depends on the various parameters including the size of the leads and the phonon distribution at the boundaries\cite{RN26}. To eliminate these extrinsic factors, we assume an interface shared by two semi-infinite leads as shown in Fig. \ref{fig:1} It is expected that phonons near the interface are significantly deviated from equilibrium due to the interfacial reflection and transmission; however, the phonon distribution far from the interface should be the same as the bulk distribution. Therefore, the semi-infinite leads can be replaced with finite leads that have the bulk distribution as boundary conditions at $x=-L_{\text{Si}}$ and $x=L_{\text{Ge}}$:
\begin{align} \label{eq:fbulk2}
f_{i}|_{x=-L_{\text{Si}}} = f_{i}^{0}(T_{\text{H}}) - v_{i,x} \tau_{i} \frac{\partial f_{i}^{0}}{\partial T} \frac{dT}{dx} \text{ for modes } i \text{ with } v_{i,x}>0
\end{align}
\begin{align} \label{eq:fbulk3}
f_{i}|_{x=L_{\text{Ge}}} = f_{i}^{0}(T_{\text{C}}) - v_{i,x} \tau_{i} \frac{\partial f_{i}^{0}}{\partial T} \frac{dT}{dx} \text{ for modes } i \text{ with } v_{i,x}<0
\end{align}
where $f^{0}$ is the Bose-Einstein distribution. Note that the above boundary conditions are the analytic solution of Eq. \ref{eq:PBE1} in an infinitely large sample under a homogenous temperature gradient.
%% figure 1

For the boundary conditions in Eqs. \ref{eq:fbulk2} and \ref{eq:fbulk3}, we assign the local equilibrium temperature values ($T_{\text{H}}$ and $T_{\text{C}}$) at the boundaries. However, four other variables ($dT/dx$ at $x=-L_{\text{Si}}$ and $x=L_{\text{Ge}}$, $L_{\text{Si}}$ and $L_{\text{Ge}}$) need to be found to satisfy the self-consistency of the boundary conditions. For example, the temperature gradients at boundaries should satisfy the Fourier’s law,  $-dT/dx=q"/\kappa_{\text{bulk}}$ where q" is the heat flux from the solution the PBE and $\kappa_{\text{bulk}}$ is the intrinsic thermal conductivity of a bulk lead. Also, the lengths of the two leads should be large enough so that the phonons leaving the computational domain follow the bulk distribution, i.e.,
\begin{align} \label{eq:fbulk4}
f_{i}|_{x=-L_{\text{Si}}} = f_{i}^{0}(T_{\text{H}}) - v_{i,x} \tau_{i} \frac{\partial f_{i}^{0}}{\partial T} \frac{dT}{dx} \text{ for modes } i \text{ with } v_{i,x}<0
\end{align}
\begin{align} \label{eq:fbulk5}
f_{i}|_{x=L_{\text{Ge}}} = f_{i}^{0}(T_{\text{C}}) - v_{i,x} \tau_{i} \frac{\partial f_{i}^{0}}{\partial T} \frac{dT}{dx} \text{ for modes } i \text{ with } v_{i,x}>0
\end{align}

Such self-consistent boundary conditions are found by solving the PBE iteratively. In the beginning, we solve the PBE assuming certain lengths of leads and zero temperature gradients at boundaries. The heat flux, $q"$, from the solution of the PBE is used to update the temperature gradient at boundaries for the next iteration. Such an iterative process of solving the PBE with the assumed lead lengths continue until heat fluxes from the two consecutive iterations have a difference less than 1\%. Then, the iterative method for solving the PBE is repeated with varying lead lengths. From a series of simulation, we found that lead is sufficiently long when the length is  200$\Lambda_{\text{avg}}$ on each side, where $\Lambda_{\text{avg}}$ is mode-averaged mean free path of phonons weighted by mode specific heat calculated as
\begin{align} \label{eq:lambda_avg6}
\Lambda_{\text{avg}} = \frac{\sum_{i} \hbar \omega_{i} \frac{\partial f_{i}^{\text{eq}}}{\partial T}\lambda_{i}}{\sum_{i} \hbar \omega_{i} \frac{\partial f_{i}^{\text{eq}}}{\partial T}} 
\end{align}
for each constituent material. Here, $\hbar$ is the reduced Planck constant, $f_i^{\text{eq}}$, $\omega_{i}$ and $\lambda_{i}$ are the equilibrium distribution at global equilibrium temperature, the frequency and mean free path of the phonon mode $i$, respectively. Detailed data can be found in the Supplementary Information (SI) 2. In our work, the lead lengths for all cases are fixed at 200$\Lambda_{\text{avg}}$.

For each iteration step, the PBE is solved stochastically using the deviational kinetic Monte Carlo (MC) method\cite{RN27}. The interfacial scattering matrix is calculated using the DMM\cite{RN28}. All other inputs such as phonon dispersion and three-phonon scattering rates are calculated from first-principles using VASP\cite{RN29,RN30,RN31,RN32,RN33,RN34}, Phonopy\cite{RN35} and ShengBTE\cite{RN36}. Details of lattice dynamic calculations and input files to the PBE solver are provided in the SI 1. Other details regarding solving the PBE using the deviational kinetic MC method are provided in the SI 2.

The phonon distribution from the MC simulation is post-processed to obtain local thermophysical properties. Each lead is divided into ten equal-sized control volumes in which the distribution function is spatially averaged to reduce the variance. The local temperature and heat flux are found from the space-averaged phonon distribution function.
\begin{align} \label{eq:localT7}
T_{\text{loc}} = \left( C_{\text{V}} N V_{\text{uc}} \right) ^{-1} \sum_{i} \hbar \omega_{i} f_{i} 
\end{align}
 where $N$ is the number of wavevectors in reciprocal space, $V_{\text{uc}}$ is the volume of unit cell, and $C_\text{V}$ is the volumetric specific heat. The heat flux $q"$ is defined as
\begin{align} \label{eq:q8}
q'' = \left( N V_{\text{uc}} \right) ^{-1} \sum_{i} v_{i,x} \hbar \omega_{i} f_{i} 
\end{align}

The total thermal resistance calculated from the MC simulation can be divided into several thermal resistance values based on their mechanisms. The total thermal resistance in the computational domain ($R_{\text{tot}}$), defined as the heat flux over the temperature difference at the two boundaries, can be divided into two: the intrinsic thermal resistance of bulk sample ($R_{\text{bulk}}$) with the same length which is the resistance if an interface does not exist and the resistance due to the interface ($R_{\text{int}}$). The $R_{\text{bulk}}$ is from the three-phonon scattering of phonons with bulk distribution and is simply found as $L_{\text{Si}}\kappa_{\text{Si}}^{-1}+L_{\text{Ge}}\kappa_{\text{Ge}}^{-1}$, where $\kappa$ is the intrinsic bulk thermal conductivity. The $R_{\text{int}}$ is obtained by subtracting $R_{\text{bulk}}$ from $R_{\text{tot}}$. The $R_{\text{int}}$ is further divided into two parts with different mechanisms: i) the aforementioned resistance due to the three-phonon scattering of phonons with excessively non-equilibrium distribution near the interface ($R_{\text{neq}}$) and ii) the resistance directly caused by the interface scattering ($R_{\text{int}}^0$).

The non-equilibrium phonons near the interface and the resulting thermal resistance are further investigated by calculating the local entropy generation due to the three-phonon scattering. The rate of local entropy generation due to scattering is calculated as\cite{RN37}
\begin{align} \label{eq:localS9}
\dot{S}(x) = - \left( T_{\text{loc}} N V_{\text{uc}} \right) ^{-1} \sum_{i} \phi_{i}(x) \dot{f}_{i}(x)
\end{align}
where $\dot{f}_{i}(x)$ is the rate of distribution change by scattering, and $\phi_{i}(x)$ is the deviation of phonon distribution from local equilibrium defined as
\begin{align} \label{eq:phi10}
f_{i} = f_{i}^{\text{loc}} - \phi_{i} \frac{\partial f_{i}^{\text{loc}}}{\hbar \partial \omega_{i}}
\end{align}
With $\dot{f}_{i}$ in the RTA being $\left(f_{i}^{\text{loc}}-f_{i}\right)/\tau_{i}$ in Eq. \ref{eq:PBE1}, Eq. \ref{eq:localS9} becomes
\begin{align} \label{eq:S11}
\dot{S} = \frac{k_{\text{B}}}{NV_{\text{uc}}} \sum_{i} \frac{(f_{i}-f_{i}^{\text{loc}})^2}{f_{i}^{\text{loc}}(f_{i}^{\text{loc}}+1)\tau_{i}}
\end{align}
which shows that the rate of entropy generation is proportional to $\sum_{i}\left(f_i-f_i^{\text{loc}}\right)^2$ imposing a challenge regarding statistical uncertainty. For temperature and heat flux, their values are proportional to $\sum_{i}\left(f_i-f_i^{\text{loc}}\right)$ where the statistical uncertainty from a mode can be cancelled by that from another mode. Thus, relatively small number of sampling particles are sufficient for achieving small statistical uncertainty of the temperature and heat flux. However, Eq. \ref{eq:S11} shows that the statistical uncertainty of entropy generation from a mode cannot be cancelled by other modes and the uncertainty only accumulates. Thus, calculation of the entropy with a reasonably small uncertainty would require extremely large number of sampling particles. This issue can be solved by noting that the Eq. \ref{eq:S11} is similar to the variance of the phonon distribution function and a variance is inversely proportional to the number of sampling particles ($N_p$). Indeed, the function  $N_p^{-1}$ fits very well the entropy generation rate in a wide range of the number of sampling particles (see SI 2.3). Therefore, we extrapolate the rate of local entropy generation with $N_p^{-1}$ in the limit of $N_p^{-1}$ approaching to zero.

The second law of thermodynamics dictates the balance between entropy flux and generation
\begin{align} \label{eq:S12}
\dot{S} = \nabla \cdot \left(\frac{\mathbf{q}''}{T}\right)
\end{align}
where $\mathbf{q}''/T$ represent entropy flux. Using $R'=\nabla\left(-T\right)/\mathbf{q}"$ and $\nabla^2T=0$ from the energy conservation, Eq. \ref{eq:S12} becomes
\begin{align} \label{eq:R13}
R' = \left(\frac{T}{q''}\right)^{2}\dot{S}
\end{align}
Equation \ref{eq:R13} can be verified for the case of thermal transport in bulk materials (see SI 2.4). In this study, we separate the local resistivity $R'$ into two components ${R'}_{\text{bulk}}$ and ${R'}_{\text{neq}}$ which represent the resistivity of bulk lead and the excessive resistivity due to the non-equilibrium phonons from the interfacial scattering, respectively. Therefore, 
\begin{align} \label{eq:R14}
R' = R'_{\text{bulk}}+R'_{\text{neq}} = \left(\frac{T}{q''}\right)^{2}(\dot{S}_{\text{bulk}}+\dot{S}_{\text{neq}})
\end{align}
Then, the $R'_{neq}$ can be simply found as
\begin{align} \label{eq:R15}
R'_{\text{neq}} = \left(\frac{T}{q''}\right)^{2}(\dot{S}-\dot{S}_{\text{bulk}})
\end{align}
The resistance from $R'_{\text{neq}}$ in an entire lead is
\begin{align} \label{eq:R16}
R_{\text{neq}} =  \int_{\text{lead}} R'_{\text{neq}}\,dx 
\end{align}

\newpage
\section{Results and discussion}
\label{Results}

\subsection{\textbf{Significant non-equilibrium in Ge}}
\label{3.1}
\begin{figure}[H]
\includegraphics[width=\textwidth]{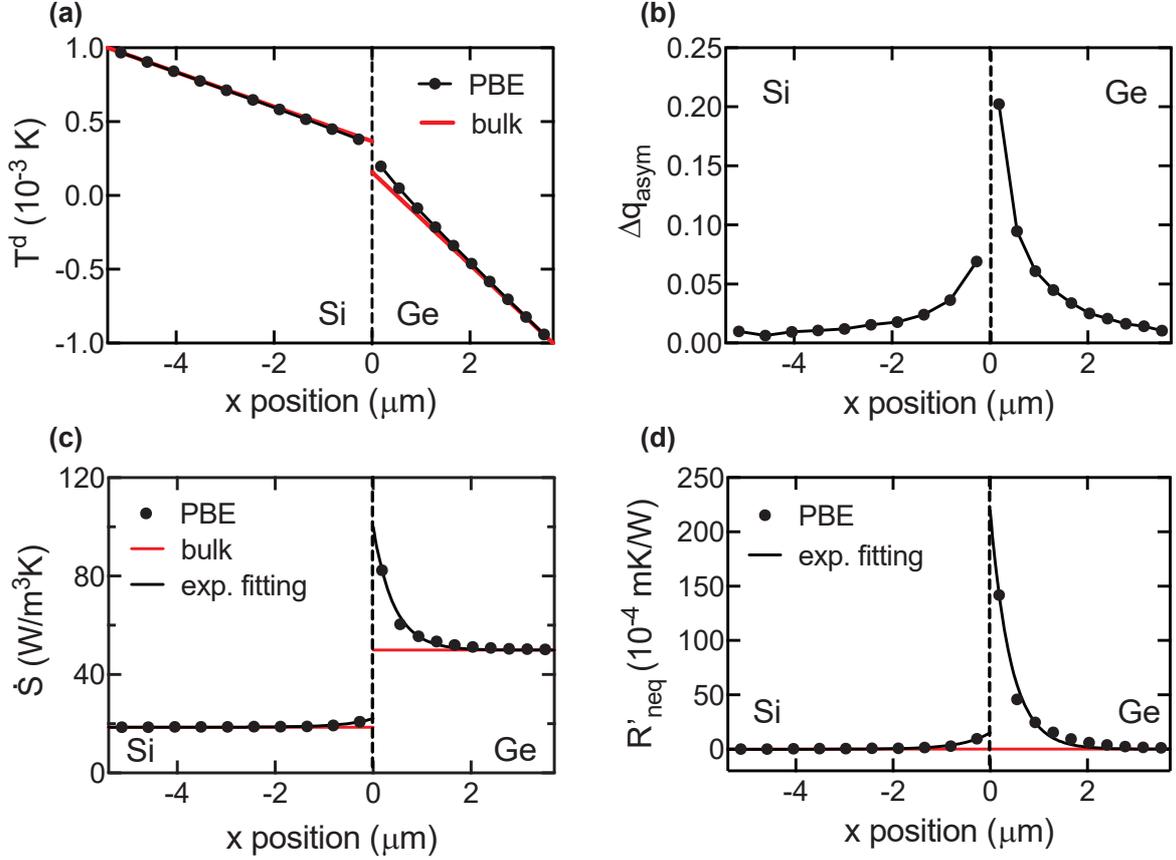}
\caption{Interfacial thermal transport across a Si-Ge interface from MC simulation of the PBE at 300 K. (a) temperature deviation ($T^\text{d}$) from 300 K showing a larger temperature gradient near the interface in Ge, (b) asymmetry of the heat flux showing large non-equilibrium of phonon distribution near the interface, (c) the rate of local entropy generation showing the excessive entropy generation in Ge, and (d) the local resistivity from the phonon-phonon scattering of non-equilibrium phonons. The deviational temperatures at boundaries are given as $\pm{0.001}$ K. The red line in (a) represents the temperature profile when the Fourier’s law is used with the intrinsic bulk thermal conductivity of Si and Ge. The black lines in (c) and (d) are exponential fittings of the black dots. The red line in (c) is the rate of local entropy generation in an infinitely large sample without an interface (${\dot{S}}_{\text{bulk}}$) and thus the area between black and red lines shows the excessive entropy generation due to the non-equilibrium phonons caused by the interface scattering. The red line in (d) is an eye-guide of $R'_{\text{neq}}=0$ for an infinitely large sample without an interface.}
\label{fig:2}
\end{figure}

Figure \ref{fig:2}(a) shows the profile of local temperature deviation, $T^\text{d}$, from reference temperature (300 K) for a Si-Ge interface. The temperature profile confirms the self-consistency of our boundary conditions. The temperature gradients far from the interface are constants and agree well with that predicted by the Fourier’s law with the intrinsic bulk thermal conductivity. Thus, the transport near the boundaries is similar to that in an infinitely large lead with the bulk phonon distribution and the finite leads are equivalent to the semi-infinite leads. Approaching the interface, the temperature profiles become non-linear and temperature gradients become larger, particularly noticeable in Ge. Such a non-linear temperature profile indicates that the thermal resistance near the interface is larger than that in the bulk region, possibly due to the resistance from the three-phonon scattering of highly non-equilibrium phonons near the interface.
%% figure 2

The degree of phonon non-equilibrium near the interface is further investigated by calculating the asymmetry of heat flux, $\Delta{q}_{\text{asym}}$, defined as $\left|q^{+}-q^{-}\right|/\left(q^{+}+q^{-}\right)$. Here, $q^+$ and $q^-$ are the heat flux contribution from phonon modes with positive and negative group velocity along the heat flow direction, respectively (see SI 2.1 for details). When phonons follow the bulk distribution, $\Delta{q}_{\text{asym}}$ is zero since two phonon modes that are related with time reversal symmetry contribute the same amount of heat flux, i.e., $\hbar\omega_iv_{i,x}^2\tau_i\left(df_i^{\text{eq}}/dx\right)$. Therefore, $\Delta{q}_{\text{asym}}$ measures the deviation from $f^{\text{bulk}}$ due to scattering mechanisms that is asymmetric in space, for example, phonon-interface scattering in this work. In Fig. 2(b), the calculated asymmetry of heat flux is the highest near the interface and decays to zero as approaching the boundaries. The asymmetry is particularly high in Ge implying that the phonon distribution in Ge significantly differs from the bulk distribution. The $\Delta{q}_{\text{asym}}$ being almost zero near the boundaries confirm that the local phonon distributions near the boundaries are very close to the bulk distribution. 

The phonons with highly non-equilibrium distribution caused by the interface scattering result in large entropy generation and thus thermal resistance upon three-phonon scattering in a lead. According to the Boltzmann’s H-theorem, any scattering process generates entropy unless the distribution is the equilibrium one, and the generated entropy is proportional to the deviation of the distribution from the equilibrium case\cite{RN37} as shown in Eq. \ref{eq:localS9} Therefore, as the highly non-equilibrium phonons near the interface diffuse away from the interface and experience three-phonon scattering, each scattering event generates entropy and contributes to the thermal resistance. In Fig. \ref{fig:2}(c), we show the local entropy generation rates ($\dot{S}$) calculated from phonon distribution and scattering rates. The red horizontal line is an eye-guide for the entropy generation rate in a bulk sample assuming the bulk distribution ($\dot{S}_{bulk}$) which results in $R_{bulk}$. The difference between black and red lines represents the excessive entropy generation from the non-equilibrium phonon distribution that the interface scattering causes. The $\dot{S}$ far from the interface is identical to $\dot{S}_{bulk}$, supporting that the distribution at the boundaries is close to the bulk distribution. We observe that the values of $\dot{S}$ near the interface are significantly larger than the bulk values, particularly in Ge, which is consistent with the large temperature gradient and asymmetry of heat flux in Figs. \ref{fig:2}(a-b). The large excessive entropy generation in Fig. \ref{fig:2}(c) exists up to 2 to 3 \si{\um} from the interface in Ge. This highlights that the effects of the interface on thermal transport are not limited to the proximity of the interface (e.g., several tens nanometer), but exist in a distance of a few micrometers from the interface. Figure \ref{fig:2}(d) shows the local thermal resistivity due to the non-equilibrium phonons ($R'_{neq}$) from the local entropy generation calculated with $\dot{S}-{\dot{S}}_{bulk}$ as in Eq. \ref{eq:R13}. Corresponding to the larger excessive entropy generation, the Ge side has larger local thermal resistivity than the Si side. In addition, it is noteworthy that Ge requires longer distance than Si for relaxing the non-equilibrium phonons to the bulk distribution even though the phonon mean free path is shorter in Ge compared to Si. This is because the degree of non-equilibrium at the interface is much more pronounced in Ge than in Si.

\begin{figure}[H]
\includegraphics[width=\textwidth]{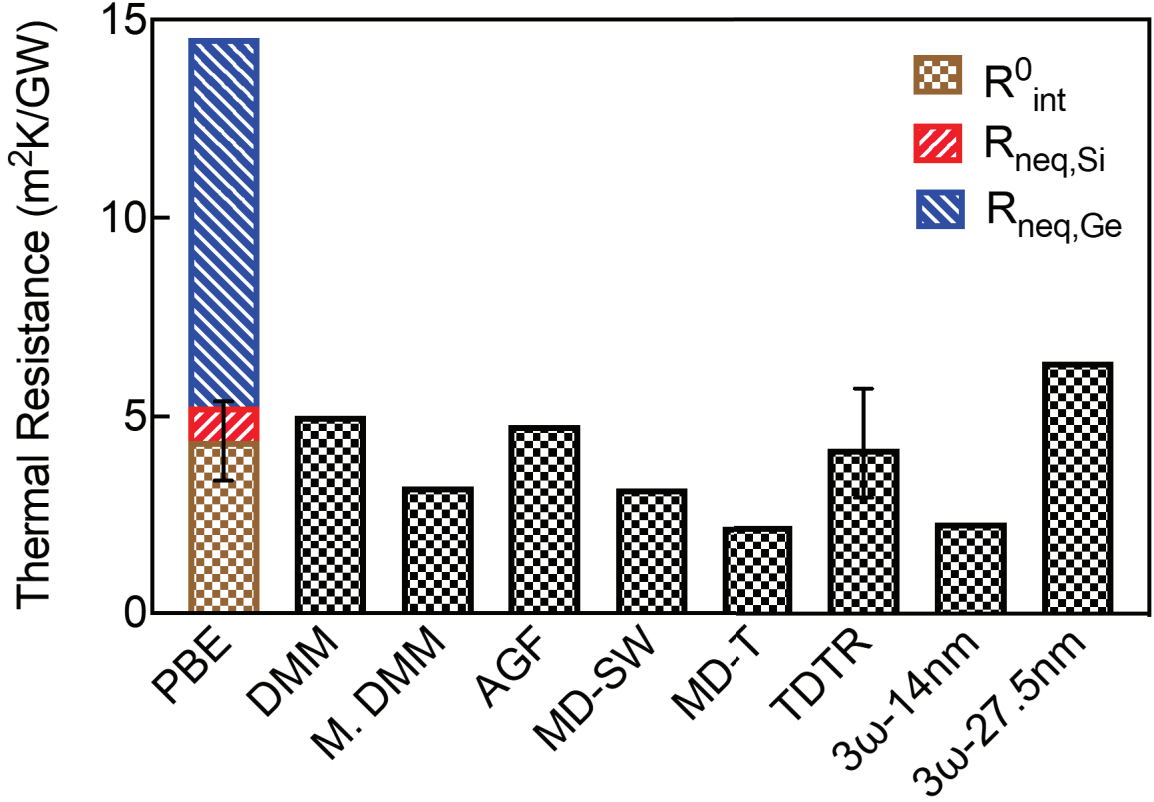}
\caption{Breakdown of the interfacial resistance at 300 K showing significant thermal resistance from the non-equilibrium phonons in Ge side and comparison with interfacial resistance from previous studies. The thermal resistance from the PBE simulation at 300 K is decomposed into the three components: interface ($R_{\text{int}}^0$), non-equilibrium contribution in Si ($R_{\text{neq},\text{Si}}$) and Ge ($R_{\text{neq},\text{Ge}}$). The DMM and M. DMM represent the diffuse mismatch model with the Landauer formalism and the modified Landauer formalism. The literature values are from the atomistic Green’s function with the Landauer formalism (AGF)\cite{RN38}, non-equilibrium molecular dynamics simulation with Stillinger-Weber potential (MD-SW)\cite{RN22} and Tersoff potential (MD-T)\cite{RN39}, the thermoreflectance measurement (TDTR) of a single Si-Ge interface\cite{RN40}, and 3$\omega$ measurements for superlattices with period thickness of 14 nm\cite{RN41} and 27.5 nm\cite{RN42}.}
\label{fig:3}
\end{figure}

With the local entropy generation rate calculated from the PBE, we decompose $R_{int}$ into three components: the resistance from the scattering of non-equilibrium phonons in Si and Ge leads separately ($R_{neq,Si}$ and $R_{neq,Ge}$) and the resistance directly caused by the interface scattering ($R_{int}^0$). The $R_{neq,Si}$ and $R_{neq,Ge}$ are the area between black and red lines in Fig. \ref{fig:2}(d). The $R_{int}^0$ is simply found by subtracting $R_{neq,Si}$ and $R_{neq,Ge}$ from $R_{int}$. The decomposed $R_{int}$ is shown in Fig. \ref{fig:3} The estimation of error bar in Fig. \ref{fig:3} is detailed in the SI 1.3. While $R_{neq,Si}$ is small, $R_{neq,Ge}$ is noticeably significant and twice larger than $R_{int}^0$. The $R_{neq,Si}$ and $R_{neq,Ge}$ contribute around 70\% to $R_{int}$, indicating that the primary interfacial resistance of the Si-Ge interface is from the non-equilibrium effect rather than the interface scattering itself. Thus, it is critical to consider coupled effects of three-phonon and phonon-interface scattering for the overall thermal interfacial resistance. 
%% figure 3

Figure \ref{fig:3} also compares the interfacial thermal resistance of this work with previous simulation and experimental studies. Overall, $R_{\text{int}}$ from the PBE simulation is much larger than resistances from the previous studies but $R_{\text{int}}^0$ is comparable to the previously reported values. It is noteworthy that our definition of the interfacial resistance, $R_{\text{int}}$, is different from the interfacial resistance of the previous studies. The interfacial resistance in the previous studies using the Landauer formalism and NEMD is defined as heat flux divided by a temperature drop at the interface and does not include the effects of non-equilibrium phonon scattering near the interface. Thus, the interfacial resistance in the previous simulation studies is similar to $R_{\text{int}}^0$. The interfacial resistance from experiments captures the non-equilibrium effect, but we expect in much less extent due to the small size of leads. The thickness of the Ge layer in the TDTR experiment is 250 nm\cite{RN40}. For the superlattices, the Si and Ge layer thickness is smaller than 7 nm\cite{RN41} and 14 nm\cite{RN42}.

\subsection{\textbf{Role of anharmonic process for interfacial thermal transport}}
\label{3.2}

We now discuss how anharmonic three-phonon scattering in the leads affect the thermal resistance with increasing temperature particularly above the Debye temperature of Ge ($\theta_{\text{D},\text{Ge}}$)  which is 371 K\cite{RN43}. In general, if the interfacial thermal transport is assumed a harmonic process, the interfacial thermal resistance shows similar trend as the inverse of specific heat of the constituent material with a lower Debye temperature; the interfacial thermal resistance decreases with temperature below the Debye temperature but becomes nearly constant above it. In the past, the interfacial thermal resistance that does not follow this trend was often observed when two constituent materials have highly mismatched Debye temperatures such as a Pb-diamond interface\cite{RN44}. Such phenomena were considered evidence of a significant anharmonic process in the interfacial thermal transport. 

\begin{figure}[H]
\includegraphics[width=\textwidth]{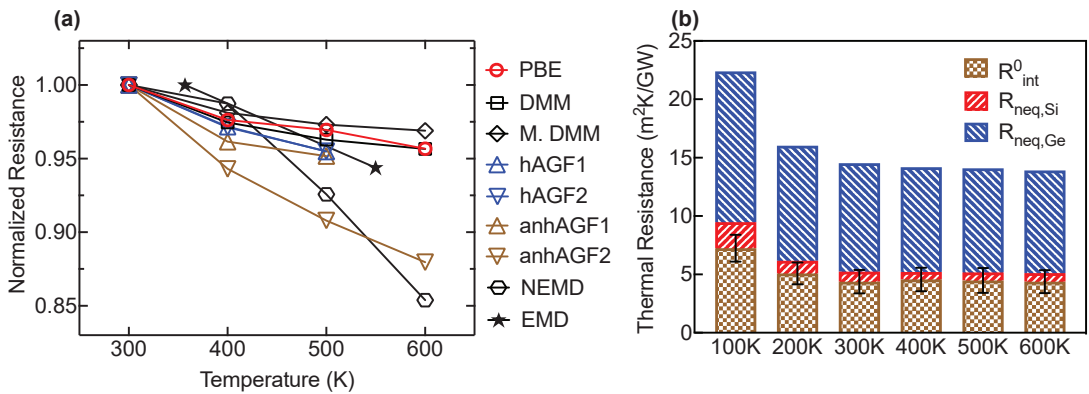}
\caption{Interfacial thermal resistance as a function of temperature. (a) Interfacial thermal resistance normalized by the values at 300 K (350K for EMD) from different methods including diffuse mismatch model with Landauer formalism (DMM) and with the modified Landauer formalism (M. DMM), harmonic AGF (hAGF1\cite{RN38} and hAGF2\cite{RN13}), anharmonic AGF (anhAGF1\cite{RN12} and anhAGF2\cite{RN13}), EMD\cite{RN45}, and NEMD\cite{RN22}. (b) Decomposed interfacial thermal resistance from the PBE simulation as a function of temperature.}
\label{fig:4}
\end{figure}

For Si-Ge interface, it has not been clear whether the interfacial resistance decreases with temperature above $\theta_{\text{D},\text{Ge}}$ since the past studies show disagreement as seen in Fig. \ref{fig:4}(a). When the Landauer formalism is used with elastic interfacial scattering model such as DMM and harmonic AGF\cite{RN38}, the interfacial resistance is nearly constant above $\theta_{\text{D},\text{Ge}}$. The two previous studies using the anharmonic AGF, which consider inelastic interfacial scattering, show opposite results from each other; one shows decreasing resistance with temperature above $\theta_{\text{D},\text{Ge}}$\cite{RN13} while the other does not\cite{RN12}. An NEMD simulation using Stillinger-Weber potential shows decreasing resistance above $\theta_{\text{D},\text{Ge}}$\cite{RN22}. However, an NEMD simulation using Tersoff potential, which is not shown in Fig. 4 since it reported only two data points at 300 and 1000 K, shows only 7\% of resistance decrease from 300K to 1000K\cite{RN17}. An equilibrium molecular dynamics (EMD) simulation with Stillinger-Weber potential for Si-Ge superlattices with a period thickness of 20 nm report temperature-independent thermal conductance above $\theta_{\text{D},\text{Ge}}$ until 1000K\cite{RN45}. 

%% figure 4

Figure \ref{fig:4}(b) presents the decomposed interfacial thermal resistance from the PBE simulation at different temperatures. All three components ($R_{\text{int}}^0$, $R_{\text{neq},\text{Si}}$ and $R_{\text{neq},\text{Ge}}$) do not show noticeable changes above $\theta_{\text{D},\text{Ge}}$, indicating that the three-phonon scattering in the leads does not reduce the interfacial resistance as temperature increases for Si-Ge interface. We need to note that our simulation includes three-phonon scattering in the leads but does not consider the inelastic interfacial scattering. Thus, if the inelastic interfacial scattering is significant in Si-Ge interface, our simulation has a limited capability in describing anharmonic process for the interfacial thermal transport. However, our result clearly shows that the three-phonon scattering in the Si and Ge leads alone cannot reduce the interfacial thermal resistance with temperature above $\theta_{\text{D},\text{Ge}}$.

\begin{figure}[H]
\includegraphics[width=\textwidth]{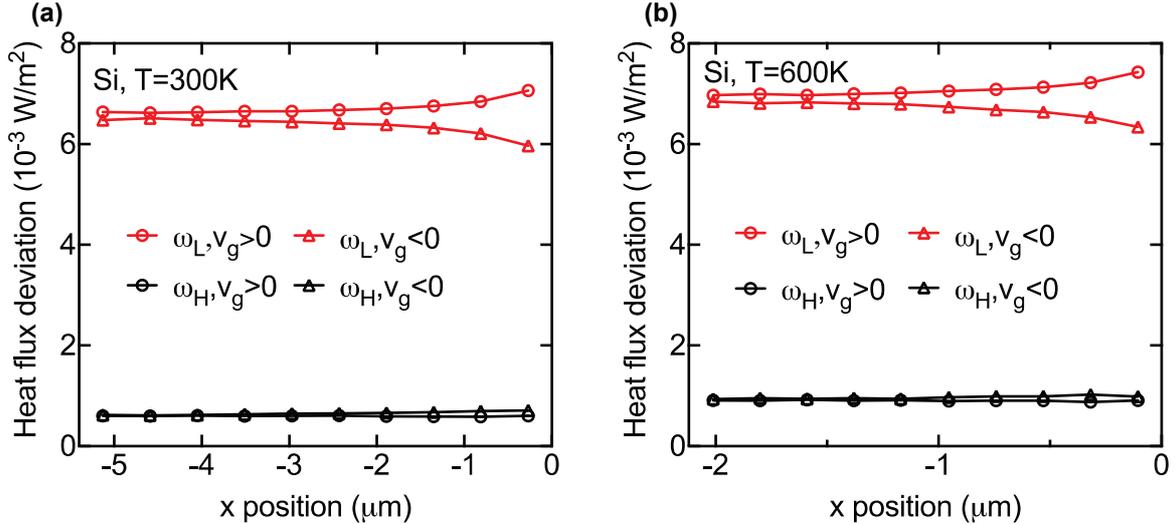}
\caption{Deviational heat flux in Si for phonon frequency below the maximum frequency of Ge phonon ($\omega_\text{L}$)  and phonon frequency above the maximum frequency of Ge phonon ($\omega_\text{H}$) at (a) T=300K and (b) T=600K.}
\label{fig:5}
\end{figure}

Our simulation results seem to have conflicts in a few aspects with previous studies\cite{RN46,RN47} that used multi-temperature models for interfacial transport with similar size leads. While our simulation shows significant non-equilibrium resistance in Ge than in Si, those two studies show more pronounced non-linear temperature profiles in Si. One of the two studies shows that the overall interfacial thermal resistance including the non-equilibrium effect is not significantly different from the resistance from the Landauer formalism\cite{RN47}; however, as shown in Fig. \ref{fig:3} and Fig. \ref{fig:4}(b), $R_{\text{neq}}$ in our case exceeds $R_{\text{int}}^0$ and the overall $R_{\text{int}}$ is much larger than that from the Landauer formalism. The other study\cite{RN46} reported the significant amount of heat carried by high frequency phonons in Si although those high frequency phonons cannot be transmitted into Ge under the assumption of elastic interface scattering. They explained that the heat carried by high frequency phonons in Si is transferred to low frequency phonons by three-phonon scattering near the interface and then the transferred energy can cross the interface. However, we do not observe such a significant energy transfer between high and low frequency phonon modes in Si. Figure \ref{fig:5} shows the local heat flux deviation $q^d$, defined as $q^d=\left(NV_{\text{uc}}\right)^{-1} \sum_{i} \hbar \omega_{i}(f_i-f_{i}^{\text{loc}})v_{i,x}$,  in Si for four different groups of phonons: high and low frequency phonons ($\omega_{\text{H}}$ and $\omega_{\text{L}}$) with $v_x>0$  and $v_x<0$. The high and low frequency represents the frequency higher and lower, respectively, than the maximum frequency of Ge phonons. Note that $q^d$ excludes the contribution from local equilibrium distribution, $f^{\text{loc}}$, since it does not contribute to the overall heat flux. From Fig. \ref{fig:5}(a), the heat carried by high frequency phonons in Si is insignificant and not much different from the bulk case that can be observed near the boundary at $x=-5.4$ \si{\um}. The same trend is observed at higher temperature 600 K in Fig. \ref{fig:5}(b). Thus, there is no significant energy exchange between the high and low frequency phonons. Rather, we observe noticeable energy exchange between phonons with $v_x>0$ and $v_x<0$ near the interface since the distribution function is highly asymmetric with respect to $v_x$ as seen in the large $\Delta{q}_{\text{asym}}$ in Fig. \ref{fig:2}(b).

%% figure 5

These differences may be because the previous studies used different boundary conditions and interfacial scattering models. In addition, we believe the multi-temperature model the previous studies employed may not be able to completely capture the non-equilibrium effect. In the multi-temperature models, a single temperature value represents the distribution of a group of phonon states. The model can be justified if intragroup scattering is much stronger than intergroup scattering such that the phonon states in the same group maintain equilibrium while phonon states from different groups have different distribution or temperatures. For interfacial thermal transport, even two phonon states with time reversal symmetry can have largely different distribution near the interface due to the asymmetric interfacial scattering as can be seen in Fig. \ref{fig:2}(b) and Fig. \ref{fig:5}.

\subsection{\textbf{Factors that determine the degree of phonon non-equilibrium}}
\label{3.3}

We aim to explain why the Ge side has much larger entropy generation than the Si side and seek for a general rule that determines $R_{\text{neq}}$. A thorough explanation for the pronounced non-equilibrium phonon in Ge compared to Si seems not straightforward since the phonon distribution near the interface is the result of the complex interplay and balance of phonon advection, three-phonon scattering, and phonon-interface scattering as governed by the PBE. Instead, we consider several fictitious Si-Ge interfaces to study the effects of the mismatch of phonon dispersion, phonon frequency, and phonon group velocity.

\begin{figure}[H]
\includegraphics[width=\textwidth]{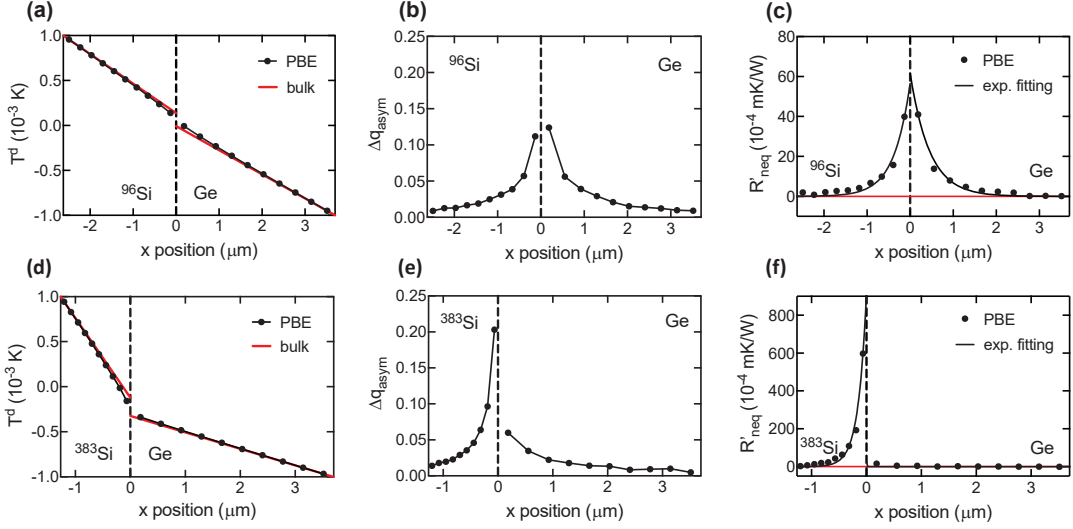}
\caption{Interfacial thermal transport across fictitious Si and Ge interfaces from the PBE simulation: (a)-(c) $^{96}$Si-Ge interface and (d)-(f) 383Si-Ge interface. (a,d), (b,e), and (c,f) are deviational temperature, asymmetry of heat flux, and local resistivity from non-equilibrium phonons, respectively.}
\label{fig:6}
\end{figure}

\begin{figure}[H]
\includegraphics[width=\textwidth]{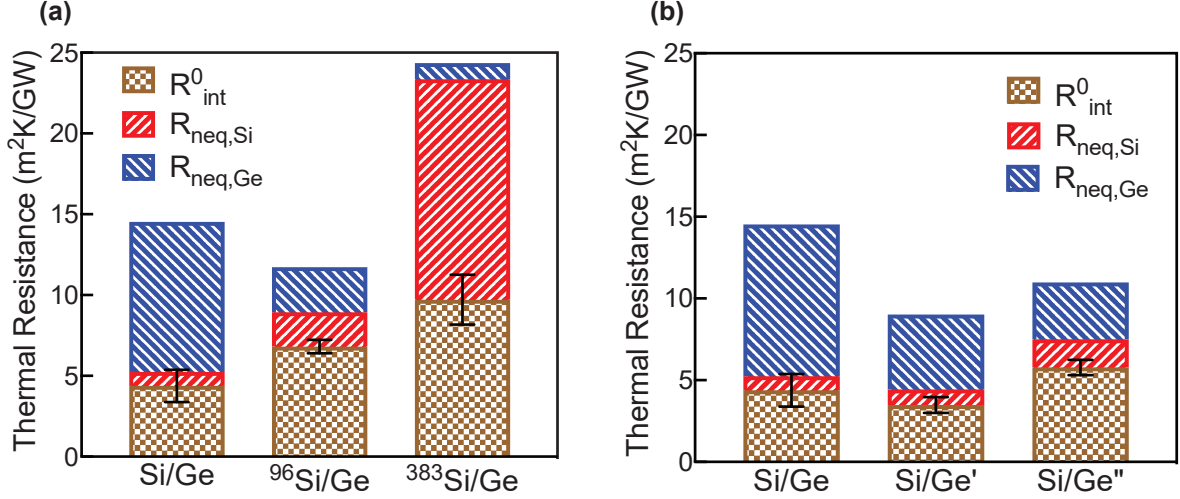}
\caption{Decomposed interfacial thermal resistance of Si-Ge interfaces at 300 K (a) with fictitious Si of which maximum frequency is the same as that of Ge ($^{96}$Si/Ge) and half of that of Ge ($^{383}$Si/Ge) and (b) with fictitious Ge which have the similar group velocity (Ge') and phonon DOS (Ge'') compared to the Si. Results of the original Si/Ge interface are shown for comparison. For all fictitious materials, the scattering rates are remained same as the original material.}
\label{fig:7}
\end{figure}

First, we consider two Si-Ge interfaces with fictitious Si leads. The original mass of Si (28 amu) is modified to 96 and 383 amu so that the maximum phonon frequency of Si is the same as and half of that of Ge, respectively. More details of comparing phonon dispersion and phonon density-of-states (DOS) of fictitious Si and Ge are presented in SI 3 and Fig. S3. The second-order force constants and three-phonon scattering rates remain the same as in the original Si case. We show the scattering rates of original Si and Ge in Fig. S4. The interfacial scattering matrix from DMM is changed because of the scaled phonon frequency of Si. The comparisons of interfacial transmissivity are shown in Figs. S5(a-c). Figure \ref{fig:6} presents the profile of deviational temperature, asymmetry of heat flux, and the local resistivity from non-equilibrium phonons for the two fictitious Si-Ge interfaces. For the $^{96}$Si-Ge interface which has a good match of phonon dispersion, we find comparable non-equilibrium effects in the two leads. The $T^d$, $\Delta{q}_{\text{asym}}$, and $R'_{\text{neq}}$ in both leads all show similar deviations from bulk cases. For the $^{383}$Si-Ge interface, however, the non-equilibrium effect is stronger in Si side than in Ge side. In Fig. \ref{fig:7}(a), we decompose the total interfacial thermal resistances of $^{96}$Si-Ge and $^{383}$Si-Ge interfaces to compare the non-equilibrium effects in each case. When the phonon dispersions are matched ($^{96}$Si and Ge), the $R_{\text{neq}}$ is similar for both sides and is smaller than dispersion mismatched cases. However, the $^{383}$Si-Ge interfaces show that the larger non-equilibrium effect is observed in the fictitious Si. This is similar to the original Si-Ge case in that the non-equilibrium effect is much larger in a lead with lower Debye temperature.

%% figure 6

%% figure 7

\begin{figure}[H]
\includegraphics[width=\textwidth]{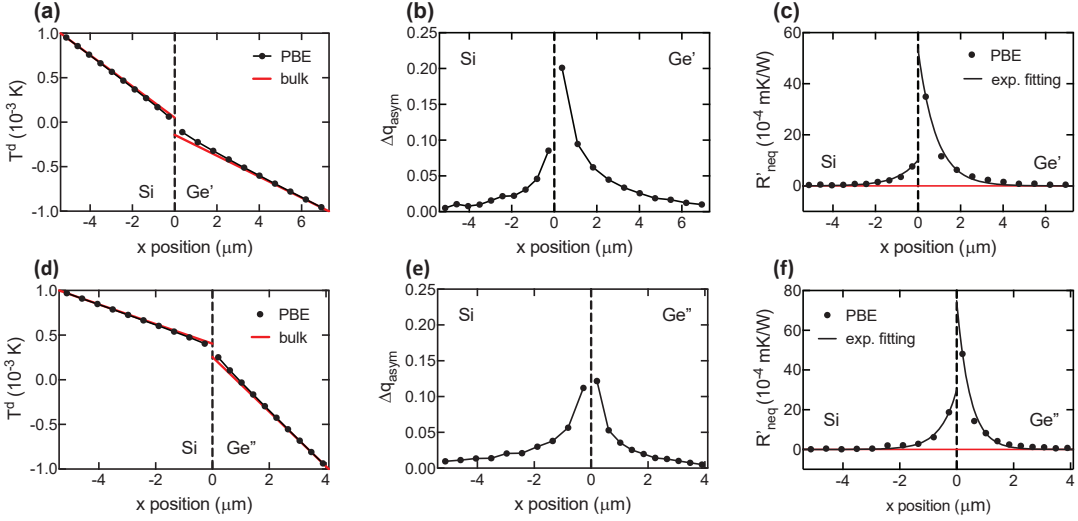}
\caption{Profiles of local deviational temperature, asymmetry of heat flux and local resistivity from non-equilibrium phonons for (a)-(c) Si-Ge' where the group velocity of the Ge is doubled for better match of group velocities between Si and Ge and (d)-(f) Si-Ge'' where the phonon DOS are matched.}
\label{fig:8}
\end{figure}

The effects of phonon dispersion mismatch are further investigated with the separate effects of phonon group velocity and phonon DOS mismatches. For the former effect, we consider a fictitious Ge lead (called Ge' in this work) where the group velocities of all phonon modes are doubled to be similar to those in Si. Note that the phonon frequencies are not changed to keep phonon DOS the same as the original Ge. We expect that a better match of group velocity would decrease the degree of non-equilibrium. This is simply because the spectral heat flux in both leads in the proximity of the interface should be the same if the interfacial scattering is elastic, i.e., $\sum_{i\in \text{Si}}{v_{i,x}\left(f_i-f_i^0\right)\delta\left(\omega_i-\omega_0\right)}=\sum_{i\in \text{Ge}}{v_{i,x}\left(f_i-f_i^0\right)\delta\left(\omega_i-\omega_0\right)}$. Thus, phonons in the lead where group velocity is lower exhibits larger deviation from the equilibrium distribution. Figures \ref{fig:8}(a-c) indeed show that the degree of phonon non-equilibrium in the Ge' lead is much reduced from that in the original Ge lead in Fig. \ref{fig:2}. Figure \ref{fig:7}(b) also shows that the ratio of $R_{\text{neq},\text{Ge}'}$ to $R_{\text{neq},\text{Si}}$ is only 4.7, much smaller than 10.6 for the original Si-Ge interface. The small group velocity of Ge compared to Si can explain why $R_{\text{neq}}$ is much more pronounced in Ge than in Si for the original Si-Ge interface. The total interfacial thermal resistance in the Si-Ge' interface is much lower than that in the Si-Ge interface, suggesting that the mismatch of group velocity plays an important role in interfacial thermal resistance. Here, we have focused on the non-equilibrium at the interface claiming that the lead with relatively lower group velocity compared to the other lead should exhibit larger degree of non-equilibrium at the interface. On the other hand, the lower group velocity may relax the non-equilibrium phonons faster with the shorter mean free path, possibly causing less $R_{\text{neq}}$. However, our solution of the PBE for the Si-Ge interfaces show that the large degree of non-equilibrium imposed at the interface of Ge side due to lower group velocity overshadows the effects of fast relaxation of non-equilibrium phonons in the Ge lead from relatively short mean free path.

%% figure 8

The effects of phonon DOS mismatch are investigated using another fictitious Ge (called Ge'' in this work) of which phonon frequency is doubled so that its phonon DOS matches well with that of Si. Note that all other properties including group velocity and scattering rates are kept the same as the original Ge. Figures \ref{fig:8}(d-f) show that the degrees of non-equilibrium in Ge'' and Si are comparable, similar to those for the 96Si-Ge shown in Figs. \ref{fig:6}(a-c). The $R_{\text{int}}$ and $R_{\text{neq}}$ for the Si-Ge'' interface (Fig. \ref{fig:7}(b)) are similar to those for the $^96$Si-Ge interface. However, $R_{\text{neq},\text{Ge}''}/R_{\text{neq},\text{Si}}$ is about 2 and different from 1.3 of $R_{\text{neq},\text{Ge}}/R_{\text{neq},{^{96}}\text{Si}}$, and the difference may be due to the mismatched group velocity in the Si-Ge'' interface. From the comparison of Si-Ge' and Si-Ge'' cases, we conclude that both DOS and group velocity mismatch are dominant factors affecting the non-equilibrium contribution to thermal resistance.

\begin{figure}[H]
\includegraphics[width=\textwidth]{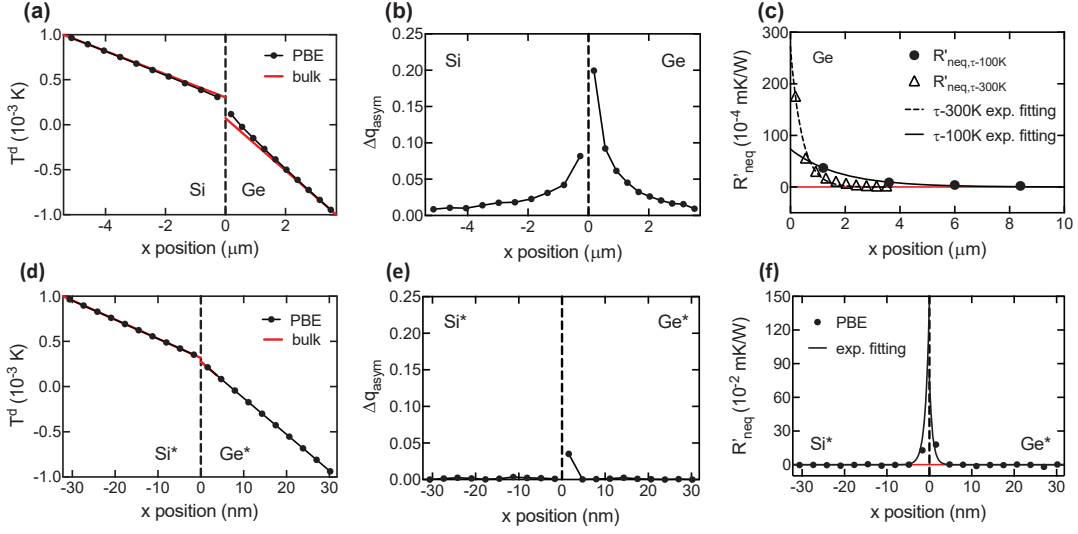}
\caption{Interfacial thermal transport for (a)-(c) Si-Ge interface at 100K with three-phonon scattering rates calculated at 300K, and (d)-(f) $\text{Si}^*$/$\text{Ge}^*$ interface with phonon relaxation time of $\tau_i=\left(2\omega_i\right)^{-1}$. In (c), the $R'_{\text{neq},\text{Ge}}$ based on the scattering rate at 300 K (noted as $\tau$-300K) is compared to the $R'_{\text{neq},\text{Ge}}$ based on the scattering rate 100 K (noted as $\tau$-100K).}
\label{fig:9}
\end{figure}

\begin{figure}[H]
\includegraphics[width=\textwidth]{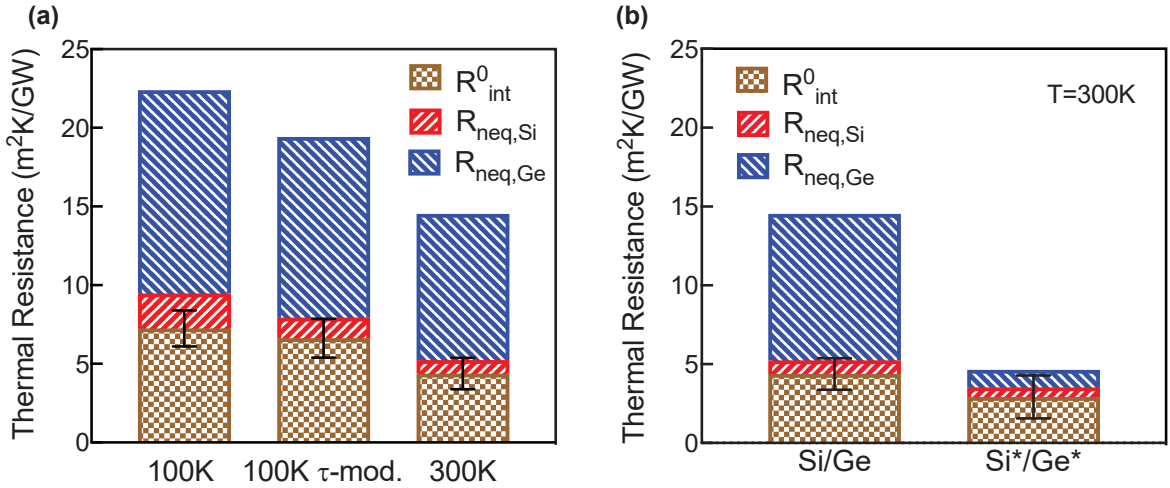}
\caption{The effects of three-phonon scattering on the $R_{\text{neq}}$ in Si-Ge interfaces. Decomposed thermal resistance for (a) a Si-Ge interface at 100 K while the three-phonon scattering rates at 300 K are applied (noted as 100K $\tau-\text{mod}$) compared to the original Si-Ge interface at 100 and 300 K (noted as 100 K and 300 K) and (b) a Si-Ge interface at 300 K with the three-phonon scattering rates of $\tau_i=\left(2\omega_i\right)^{-1}$ ($\text{Si}^*$/$\text{Ge}^*$) compared to the original Si-Ge interface at 300 K (Si/Ge).}
\label{fig:10}
\end{figure}

Lastly, we examine the effects of phonon anharmonicity on $R_{\text{neq}}$. We expect that the strong three-phonon scattering would relax the non-equilibrium phonons near the interface quickly and thereby $R_{\text{neq}}$ can be reduced. We consider a Si-Ge interface at 100 K with two different three-phonon scattering rates; one is the original scattering rates of Si and Ge at 100 K and the other is scattering rates at 300 K. Figures \ref{fig:9}(a-c) show the results of simulation at 100 K with modified scattering rates. Figures \ref{fig:9}(a-b) show similar $T^d$ and $\Delta{q}_{\text{asym}}$ as in Fig. \ref{fig:2}. In Fig. \ref{fig:9}(c), when the scattering rates at 300 K are used, $R'_{\text{neq}}$ of Ge is much increased near the interface since the scattering rate is larger. However, $R'_{\text{neq}}$ quickly decays in the space because of the fast relaxation to the bulk phonon distribution. As a result, Fig. \ref{fig:10}(a) shows the reduction of $R_{\text{neq},\text{Ge}}$ when the scattering rate at 300 K is assumed. We further increase the scattering rates by assigning the relaxation time as half of the reciprocal of modal frequency, known to lead to the lower limit of thermal conductivity\cite{RN48} , while the harmonic phonon properties of Si and Ge remain the same. This case is denoted as the Si$^*$-Ge$^*$ interface. In Figs. \ref{fig:9}(d-f), the non-equilibrium effects are significantly suppressed in both sides with such an extremely high three-phonon scattering rates. As a result, the $R_{\text{int}}$ of Si$^*$-Ge$^*$ interface is 4.64 m$^2$K/GW while that of original Si-Ge interface is 14.53 m$^2$K/GW. Figure \ref{fig:10}(b) shows that the large reduction of interfacial thermal resistance comes primarily from the reduction of $R_{\text{neq},\text{Ge}}$. Our results suggest that increasing phonon anharmonicity could reduce the interfacial thermal resistance. However, such a reduction is at cost of increased $R_{\text{bulk}}$, making the total thermal resistance, $R_{\text{tot}}=R_{\text{int}}+R_{\text{bulk}}$, much larger if the increase of $R_{\text{bulk}}$ overshadows the reduction of $R_{\text{int}}$. The $R_{\text{bulk}}$ of Si$^*$ and Ge* are 74.03 m$^2$K/GW and 138.76 m$^2$K/GW, around 45 times larger than $R_{\text{int}}$. Therefore, the total interfacial resistance $R_{\text{tot}}$ can be reduced by proper engineering of phonon scattering near an interface that reduce $R_{\text{neq}}$, but such efforts should be judicious to minimize the increase of $R_{\text{bulk}}$.

%% figure 9

%% figure 10

\newpage
\section{Conclusions}
\label{Conclusions}
In summary, we study phonon transport across a Si-Ge interface by solving the PBE in both reciprocal and real spaces using a kinetic MC method with ab initio inputs. Significant non-equilibrium phonon distribution is observed in Ge lead through asymmetry of heat flux and local entropy generation existing up to 3 micrometers away from the interface, which has not been reported in previous studies. The non-equilibrium phonons caused by the interface is relaxed to the bulk phonon distribution through three-phonon scattering and significant amount of entropy and thermal resistance is generated during the process. As a result, the non-equilibrium effect contributes to more than half of the overall interfacial thermal resistance and even larger than the resistance directly caused by the interface scattering. Simulations using several fictitious Si-Ge where phonon DOS and group velocity are modified show that non-equilibrium phonon distribution is pronounced in the lead with lower group velocity and phonon frequency scale. It also shows that the non-equilibrium thermal resistance is minimal when the phonon DOS and group velocity of two constituent materials are well matched. The non-equilibrium thermal resistance can be reduced by increasing three-phonon scattering rate. As the three-phonon scattering rate increases, the non-equilibrium distribution caused by the interface scattering is quickly relaxed to the bulk distribution reducing the non-equilibrium thermal resistance in the entire semi-infinite lead. However, such a strong three-phonon scattering also increases the bulk resistance, leading to the increase of the total resistance. Our work clearly shows that the interfacial thermal transport needs to be understood in microscale in addition to the previously studied atomistic scale; the relaxation of the non-equilibrium phonons through three-phonon scattering is a dominant contributor to the interfacial resistance for Si-Ge interface and such relaxation process takes several micrometers which corresponds to the mean free paths of thermal acoustic phonons with long wavelengths.

\newpage

%%\printbibliography
\bibliographystyle{elsarticle-num}
\bibliography{main_ref}

\begin{thebibliography}{10}
\expandafter\ifx\csname url\endcsname\relax
  \def\url#1{\texttt{#1}}\fi
\expandafter\ifx\csname urlprefix\endcsname\relax\def\urlprefix{URL }\fi
\expandafter\ifx\csname href\endcsname\relax
  \def\href#1#2{#2} \def\path#1{#1}\fi

\bibitem{RN1}
D.~G. Cahill, P.~V. Braun, G.~Chen, D.~R. Clarke, S.~Fan, K.~E. Goodson,
  P.~Keblinski, W.~P. King, G.~D. Mahan, A.~Majumdar, H.~J. Maris, S.~R.
  Phillpot, E.~Pop, L.~Shi, Nanoscale thermal transport. ii. 2003–2012,
  Applied Physics Reviews 1~(1) (2014).
\newblock \href {https://doi.org/10.1063/1.4832615}
  {\path{doi:10.1063/1.4832615}}.

\bibitem{RN2}
D.~G. Cahill, W.~K. Ford, K.~E. Goodson, G.~D. Mahan, A.~Majumdar, H.~J. Maris,
  R.~Merlin, S.~R. Phillpot, Nanoscale thermal transport, Journal of Applied
  Physics 93~(2) (2003) 793--818.
\newblock \href {https://doi.org/10.1063/1.1524305}
  {\path{doi:10.1063/1.1524305}}.

\bibitem{RN3}
A.~Giri, P.~E. Hopkins, A review of experimental and computational advances in
  thermal boundary conductance and nanoscale thermal transport across solid
  interfaces, Advanced Functional Materials 30~(8) (2019).
\newblock \href {https://doi.org/10.1002/adfm.201903857}
  {\path{doi:10.1002/adfm.201903857}}.

\bibitem{RN4}
C.~Monachon, L.~Weber, C.~Dames, Thermal boundary conductance: A materials
  science perspective, Annual Review of Materials Research 46~(1) (2016)
  433--463.
\newblock \href {https://doi.org/10.1146/annurev-matsci-070115-031719}
  {\path{doi:10.1146/annurev-matsci-070115-031719}}.

\bibitem{RN5}
E.~Pop, Energy dissipation and transport in nanoscale devices, Nano Research
  3~(3) (2010) 147--169.
\newblock \href {https://doi.org/10.1007/s12274-010-1019-z}
  {\path{doi:10.1007/s12274-010-1019-z}}.

\bibitem{RN6}
J.~Chen, X.~Xu, J.~Zhou, B.~Li, Interfacial thermal resistance: Past, present,
  and future, Reviews of Modern Physics 94~(2) (2022).
\newblock \href {https://doi.org/10.1103/RevModPhys.94.025002}
  {\path{doi:10.1103/RevModPhys.94.025002}}.

\bibitem{RN7}
N.~Mingo, L.~Yang, Phonon transport in nanowires coated with an amorphous
  material: An atomistic green’s function approach, Physical Review B 68~(24)
  (2003).
\newblock \href {https://doi.org/10.1103/PhysRevB.68.245406}
  {\path{doi:10.1103/PhysRevB.68.245406}}.

\bibitem{RN8}
N.~Mingo, L.~Yang, Erratum: Phonon transport in nanowires coated with an
  amorphous material: An atomistic green’s function approach [phys. rev. b68,
  245406 (2003)], Physical Review B 70~(24) (2004).
\newblock \href {https://doi.org/10.1103/PhysRevB.70.249901}
  {\path{doi:10.1103/PhysRevB.70.249901}}.

\bibitem{RN9}
S.~Sadasivam, Y.~Che, Z.~Huang, L.~Chen, S.~Kumar, T.~S. Fisher, The atomistic
  green's function method for interfacial phonon transport, Annual Review of
  Heat Transfer 17 (2014).

\bibitem{RN10}
J.~S. Wang, J.~Wang, J.~T. Lü, Quantum thermal transport in nanostructures,
  The European Physical Journal B 62~(4) (2008) 381--404.
\newblock \href {https://doi.org/10.1140/epjb/e2008-00195-8}
  {\path{doi:10.1140/epjb/e2008-00195-8}}.

\bibitem{RN11}
B.~Latour, N.~Shulumba, A.~J. Minnich, Ab initio study of mode-resolved phonon
  transmission at si/ge interfaces using atomistic green's functions, Physical
  Review B 96~(10) (2017) 104310.
\newblock \href {https://doi.org/10.1103/PhysRevB.96.104310}
  {\path{doi:10.1103/PhysRevB.96.104310}}.

\bibitem{RN12}
J.~Dai, Z.~Tian, Rigorous formalism of anharmonic atomistic green's function
  for three-dimensional interfaces, Physical Review B 101~(4) (2020).
\newblock \href {https://doi.org/10.1103/PhysRevB.101.041301}
  {\path{doi:10.1103/PhysRevB.101.041301}}.

\bibitem{RN13}
Y.~Guo, Z.~Zhang, M.~Bescond, S.~Xiong, M.~Nomura, S.~Volz, Anharmonic
  phonon-phonon scattering at the interface between two solids by
  nonequilibrium green's function formalism, Physical Review B 103~(17) (2021).
\newblock \href {https://doi.org/10.1103/PhysRevB.103.174306}
  {\path{doi:10.1103/PhysRevB.103.174306}}.

\bibitem{RN14}
N.~Mingo, Anharmonic phonon flow through molecular-sized junctions, Physical
  Review B 74~(12) (2006).
\newblock \href {https://doi.org/10.1103/PhysRevB.74.125402}
  {\path{doi:10.1103/PhysRevB.74.125402}}.

\bibitem{RN15}
K.~Gordiz, A.~Henry, Phonon transport at interfaces: Determining the correct
  modes of vibration, Journal of Applied Physics 119~(1) (2016).
\newblock \href {https://doi.org/10.1063/1.4939207}
  {\path{doi:10.1063/1.4939207}}.

\bibitem{RN16}
K.~Gordiz, A.~Henry, \href{https://www.ncbi.nlm.nih.gov/pubmed/26979787}{Phonon
  transport at crystalline si/ge interfaces: The role of interfacial modes of
  vibration}, Sci Rep 6 (2016) 23139.
\newblock \href {https://doi.org/10.1038/srep23139}
  {\path{doi:10.1038/srep23139}}.
\newline\urlprefix\url{https://www.ncbi.nlm.nih.gov/pubmed/26979787}

\bibitem{RN17}
T.~Feng, Y.~Zhong, J.~Shi, X.~Ruan, Unexpected high inelastic phonon transport
  across solid-solid interface: Modal nonequilibrium molecular dynamics
  simulations and landauer analysis, Physical Review B 99~(4) (2019).
\newblock \href {https://doi.org/10.1103/PhysRevB.99.045301}
  {\path{doi:10.1103/PhysRevB.99.045301}}.

\bibitem{RN18}
T.~Feng, W.~Yao, Z.~Wang, J.~Shi, C.~Li, B.~Cao, X.~Ruan, Spectral analysis of
  nonequilibrium molecular dynamics: Spectral phonon temperature and local
  nonequilibrium in thin films and across interfaces, Physical Review B 95~(19)
  (2017).
\newblock \href {https://doi.org/10.1103/PhysRevB.95.195202}
  {\path{doi:10.1103/PhysRevB.95.195202}}.

\bibitem{RN19}
Y.~Imry, R.~Landauer, \href{https://link.aps.org/doi/10.1103/RevModPhys.71.S306
  https://journals.aps.org/rmp/pdf/10.1103/RevModPhys.71.S306}{Conductance
  viewed as transmission}, Reviews of Modern Physics 71~(2) (1999) S306--S312.
\newblock \href {https://doi.org/10.1103/RevModPhys.71.S306}
  {\path{doi:10.1103/RevModPhys.71.S306}}.
\newline\urlprefix\url{https://link.aps.org/doi/10.1103/RevModPhys.71.S306
  https://journals.aps.org/rmp/pdf/10.1103/RevModPhys.71.S306}

\bibitem{RN20}
S.~Simons, \href{http://dx.doi.org/10.1088/0022-3719/7/22/009
  https://iopscience.iop.org/article/10.1088/0022-3719/7/22/009/pdf}{On the
  thermal contact resistance between insulators}, Journal of Physics C: Solid
  State Physics 7~(22) (1974) 4048--4052.
\newblock \href {https://doi.org/10.1088/0022-3719/7/22/009}
  {\path{doi:10.1088/0022-3719/7/22/009}}.
\newline\urlprefix\url{http://dx.doi.org/10.1088/0022-3719/7/22/009
  https://iopscience.iop.org/article/10.1088/0022-3719/7/22/009/pdf}

\bibitem{RN21}
G.~Chen, Diffusion–transmission interface condition for electron and phonon
  transport, Applied Physics Letters 82~(6) (2003) 991--993.
\newblock \href {https://doi.org/10.1063/1.1543239}
  {\path{doi:10.1063/1.1543239}}.

\bibitem{RN22}
E.~S. Landry, A.~J.~H. McGaughey, Thermal boundary resistance predictions from
  molecular dynamics simulations and theoretical calculations, Physical Review
  B 80~(16) (2009).
\newblock \href {https://doi.org/10.1103/PhysRevB.80.165304}
  {\path{doi:10.1103/PhysRevB.80.165304}}.

\bibitem{RN23}
S.~Merabia, K.~Termentzidis, Thermal conductance at the interface between
  crystals using equilibrium and nonequilibrium molecular dynamics, Physical
  Review B 86~(9) (2012).
\newblock \href {https://doi.org/10.1103/PhysRevB.86.094303}
  {\path{doi:10.1103/PhysRevB.86.094303}}.

\bibitem{RN24}
S.~Lee, X.~Li, R.~Guo, Thermal resistance by transition between collective and
  non-collective phonon flows in graphitic materials, Nanoscale and Microscale
  Thermophysical Engineering 23~(3) (2019) 247--258.
\newblock \href {https://doi.org/10.1080/15567265.2019.1575497}
  {\path{doi:10.1080/15567265.2019.1575497}}.

\bibitem{RN25}
S.~Lee, X.~Li, Hydrodynamic phonon transport: past, present and prospects, IOP
  Publishing, 2020, book section~10, pp. 1--26.

\bibitem{RN26}
G.~Chen, Thermal conductivity and ballistic-phonon transport in the cross-plane
  direction of superlattices, Physical Review B 57~(23) (1998) 14958.

\bibitem{RN27}
J.-P.~M. Péraud, N.~G. Hadjiconstantinou, An alternative approach to efficient
  simulation of micro/nanoscale phonon transport, Applied Physics Letters
  101~(15) (2012).
\newblock \href {https://doi.org/10.1063/1.4757607}
  {\path{doi:10.1063/1.4757607}}.

\bibitem{RN28}
E.~T. Swartz, R.~O. Pohl, Thermal boundary resistance, Reviews of Modern
  Physics 61~(3) (1989) 605--668.
\newblock \href {https://doi.org/10.1103/RevModPhys.61.605}
  {\path{doi:10.1103/RevModPhys.61.605}}.

\bibitem{RN29}
G.~Kresse, J.~Hafner, \href{https://www.ncbi.nlm.nih.gov/pubmed/10004490}{Ab
  initio molecular dynamics for liquid metals}, Phys Rev B Condens Matter
  47~(1) (1993) 558--561.
\newblock \href {https://doi.org/10.1103/physrevb.47.558}
  {\path{doi:10.1103/physrevb.47.558}}.
\newline\urlprefix\url{https://www.ncbi.nlm.nih.gov/pubmed/10004490}

\bibitem{RN30}
G.~Kresse, J.~Hafner, \href{https://www.ncbi.nlm.nih.gov/pubmed/10010505}{Ab
  initio molecular-dynamics simulation of the
  liquid-metal-amorphous-semiconductor transition in germanium}, Phys Rev B
  Condens Matter 49~(20) (1994) 14251--14269.
\newblock \href {https://doi.org/10.1103/physrevb.49.14251}
  {\path{doi:10.1103/physrevb.49.14251}}.
\newline\urlprefix\url{https://www.ncbi.nlm.nih.gov/pubmed/10010505}

\bibitem{RN31}
G.~Kresse, J.~Furthmüller, Efficiency of ab-initio total energy calculations
  for metals and semiconductors using a plane-wave basis set, Computational
  materials science 6~(1) (1996) 15--50.

\bibitem{RN32}
G.~Kresse, J.~Furthmüller, Efficient iterative schemes for ab initio
  total-energy calculations using a plane-wave basis set, Physical review B
  54~(16) (1996) 11169.

\bibitem{RN33}
G.~Kresse, J.~Furthmüller, J.~Hafner,
  \href{https://www.ncbi.nlm.nih.gov/pubmed/9975508}{Theory of the crystal
  structures of selenium and tellurium: The effect of generalized-gradient
  corrections to the local-density approximation}, Phys Rev B Condens Matter
  50~(18) (1994) 13181--13185.
\newblock \href {https://doi.org/10.1103/physrevb.50.13181}
  {\path{doi:10.1103/physrevb.50.13181}}.
\newline\urlprefix\url{https://www.ncbi.nlm.nih.gov/pubmed/9975508}

\bibitem{RN34}
G.~Kresse, D.~Joubert, From ultrasoft pseudopotentials to the projector
  augmented-wave method, Physical review b 59~(3) (1999) 1758.

\bibitem{RN35}
A.~Togo, I.~Tanaka, First principles phonon calculations in materials science,
  Scripta Materialia 108 (2015) 1--5.
\newblock \href {https://doi.org/10.1016/j.scriptamat.2015.07.021}
  {\path{doi:10.1016/j.scriptamat.2015.07.021}}.

\bibitem{RN36}
W.~Li, J.~Carrete, N.~A.~Katcho, N.~Mingo, Shengbte: A solver of the boltzmann
  transport equation for phonons, Computer Physics Communications 185~(6)
  (2014) 1747--1758.
\newblock \href {https://doi.org/10.1016/j.cpc.2014.02.015}
  {\path{doi:10.1016/j.cpc.2014.02.015}}.

\bibitem{RN37}
J.~M. Ziman, Electrons and phonons: the theory of transport phenomena in
  solids, Oxford university press, 2001.

\bibitem{RN38}
Z.~Tian, K.~Esfarjani, G.~Chen, Enhancing phonon transmission across a si/ge
  interface by atomic roughness: First-principles study with the green's
  function method, Physical Review B 86~(23) (2012).
\newblock \href {https://doi.org/10.1103/PhysRevB.86.235304}
  {\path{doi:10.1103/PhysRevB.86.235304}}.

\bibitem{RN39}
K.~Gordiz, A.~Henry, Phonon transport at interfaces between different phases of
  silicon and germanium, Journal of Applied Physics 121~(2) (2017).
\newblock \href {https://doi.org/10.1063/1.4973573}
  {\path{doi:10.1063/1.4973573}}.

\bibitem{RN40}
Z.~Cheng, R.~Li, X.~Yan, G.~Jernigan, J.~Shi, M.~E. Liao, N.~J. Hines, C.~A.
  Gadre, J.~C. Idrobo, E.~Lee, K.~D. Hobart, M.~S. Goorsky, X.~Pan, T.~Luo,
  S.~Graham, \href{https://www.ncbi.nlm.nih.gov/pubmed/34824284}{Experimental
  observation of localized interfacial phonon modes}, Nat Commun 12~(1) (2021)
  6901.
\newblock \href {https://doi.org/10.1038/s41467-021-27250-3}
  {\path{doi:10.1038/s41467-021-27250-3}}.
\newline\urlprefix\url{https://www.ncbi.nlm.nih.gov/pubmed/34824284}

\bibitem{RN41}
T.~Borca-Tasciuc, W.~Liu, J.~Liu, T.~Zeng, D.~W. Song, C.~D. Moore, G.~Chen,
  K.~L. Wang, M.~S. Goorsky, T.~Radetic, R.~Gronsky, T.~Koga, M.~S.
  Dresselhaus,
  \href{https://www.sciencedirect.com/science/article/pii/S0749603600909005}{Thermal
  conductivity of symmetrically strained si/ge superlattices}, Superlattices
  and Microstructures 28~(3) (2000) 199--206.
\newblock \href {https://doi.org/https://doi.org/10.1006/spmi.2000.0900}
  {\path{doi:https://doi.org/10.1006/spmi.2000.0900}}.
\newline\urlprefix\url{https://www.sciencedirect.com/science/article/pii/S0749603600909005}

\bibitem{RN42}
S.~M. Lee, D.~G. Cahill, R.~Venkatasubramanian, Thermal conductivity of si–ge
  superlattices, Applied Physics Letters 70~(22) (1997) 2957--2959.
\newblock \href {https://doi.org/10.1063/1.118755}
  {\path{doi:10.1063/1.118755}}.

\bibitem{RN43}
C.~A. Bryant, P.~H. Keesom,
  \href{https://link.aps.org/doi/10.1103/PhysRev.124.698}{Low-temperature
  specific heat of germanium}, Physical Review 124~(3) (1961) 698--700.
\newblock \href {https://doi.org/10.1103/PhysRev.124.698}
  {\path{doi:10.1103/PhysRev.124.698}}.
\newline\urlprefix\url{https://link.aps.org/doi/10.1103/PhysRev.124.698}

\bibitem{RN44}
H.-K. Lyeo, D.~G. Cahill, Thermal conductance of interfaces between highly
  dissimilar materials, Physical Review B 73~(14) (2006).
\newblock \href {https://doi.org/10.1103/PhysRevB.73.144301}
  {\path{doi:10.1103/PhysRevB.73.144301}}.

\bibitem{RN45}
Y.~Chalopin, K.~Esfarjani, A.~Henry, S.~Volz, G.~Chen, Thermal interface
  conductance in si/ge superlattices by equilibrium molecular dynamics,
  Physical Review B 85~(19) (2012).
\newblock \href {https://doi.org/10.1103/PhysRevB.85.195302}
  {\path{doi:10.1103/PhysRevB.85.195302}}.

\bibitem{RN46}
J.~Maassen, V.~Askarpour, Phonon transport across a si–ge interface: The role
  of inelastic bulk scattering, APL Materials 7~(1) (2019).
\newblock \href {https://doi.org/10.1063/1.5051538}
  {\path{doi:10.1063/1.5051538}}.

\bibitem{RN47}
Z.~Lu, J.~Shi, X.~Ruan, Role of phonon coupling and non-equilibrium near the
  interface to interfacial thermal resistance: The multi-temperature model and
  thermal circuit, Journal of Applied Physics 125~(8) (2019).
\newblock \href {https://doi.org/10.1063/1.5082526}
  {\path{doi:10.1063/1.5082526}}.

\bibitem{RN48}
D.~G. Cahill, S.~K. Watson, R.~O. Pohl,
  \href{https://www.ncbi.nlm.nih.gov/pubmed/10002297}{Lower limit to the
  thermal conductivity of disordered crystals}, Phys Rev B Condens Matter
  46~(10) (1992) 6131--6140.
\newblock \href {https://doi.org/10.1103/physrevb.46.6131}
  {\path{doi:10.1103/physrevb.46.6131}}.
\newline\urlprefix\url{https://www.ncbi.nlm.nih.gov/pubmed/10002297}

\end{thebibliography}

\end{document}

\begin{frontmatter}



\title{Supplementary Information\\}
\title{Thermal resistance from non-equilibrium phonons at Si-Ge interface}


\address[1]{Department of Mechanical Engineering and Materials Science, University of Pittsburgh, Pittsburgh, PA 15261, USA}
\address[2]{Department of Physics and Astronomy, University of Pittsburgh, Pittsburgh, PA 15261, USA}
\fntext[ref1]{These authors contributed equally to this work}
\fntext[ref2]{current address: Materials Science and Technology Division, Oak Ridge National Laboratory, Oak Ridge, TN 37831, USA}

\author[1]{Xun Li\fnref{ref1,ref2}}
\author[1]{Jinchen Han\fnref{ref1}}
\author[1,2]{Sangyeop Lee\corref{cor1}}
\cortext[cor1]{sylee@pitt.edu}

\end{frontmatter}


\section{Details of lattice dynamics calculations and input files preparation}
\label{S1}

Harmonic and anharmonic lattice dynamics calculations were performed using the ShengBTE package[1] with crystal structures and interatomic force constants from first principles using VASP[2-7] and Phonopy[8]. The reciprocal space was sampled with a wavevector mesh of 15×15×15 for both Si and Ge. Adaptive Gaussian functions were used to approximate the Dirac delta functions for energy conservation[1]. Phonon properties for each phonon state, including phonon wavevector, group velocity, polarization, frequency, and scattering rate, were explicitly calculated and used as inputs for the deviational kinetic Monte Carlo simulation.

\section{Details of deviational kinetic Monte Carlo simulation of the Peierls-Boltzmann transport equation}
\label{S2}

We employ the deviational energy-based kinetic Monte Carlo simulation to understand the phonon transport across an interface of two semi-infinite leads. There are several advantages of the deviational energy-based kinetic Monte Carlo simulation. First, since it samples only deviations of distributions from the global equilibrium, the variance is significantly reduced[9]. The energy-based simulation uses $e_i=\hbar\omega_if_i$ for distribution function and each sampling particle represents the same amount of energy. Thus, simply conserving the number of particles can strictly enforces the energy conservation upon scattering. In the kinetic Monte Carlo simulation, the time domain is not discretized, and the steady state can be sampled without initializing the system to achieve the steady state[10].

With the relaxation time approximation (RTA), each particle is independent to each other in the simulation. The particle is emitted from boundaries following bulk distribution and flies in the real space with group velocity undergoing phonon-phonon scattering or interface scattering until it leaves the computational domain. We assign $N_\text{p}$ particles emitted from the hot and cold boundaries with temperature $T_1$ and $T_2$ and linear temperature gradient $\left(dT/dx\right)_1$ and $\left(dT/dx\right)_2$ where the subscripts 1 and 2 represent left and right boundaries. The rates of energy emitted from the boundaries are determined using the boundary conditions in Eqs. (2) and (3) in the main text:
\begin{align} \label{eq:S1}
\dot{\epsilon}_1 = \frac{1}{N_1 V_{\text{uc},1}} \sum_{i,v_{i,x}>0} \frac{\partial e_{i}^{\text{eq}}}{\partial T} v_{i,x} \left[ 
(T_1-T_{\text{eq}})+(- \frac{dT}{dx})_1 \tau_i v_{i,x} \right]
\end{align}
\begin{align} \label{eq:S2}
\dot{\epsilon}_2 = \frac{1}{N_2 V_{\text{uc},2}} \sum_{i,v_{i,x}<0} \frac{\partial e_{i}^{\text{eq}}}{\partial T} v_{i,x} \left[ 
(T_2-T_{\text{eq}})+(- \frac{dT}{dx})_2 \tau_i v_{i,x} \right]
\end{align}
Here, $N$ is the number of wavevectors in reciprocal space, $V_{\text{uc}}$ is the volume of unit cell, $e^{\text{eq}}$ is equilibrium energy distribution at a global equilibrium temperature, $\tau^{-1}$ is the three-phonon scattering rate, and $v_x$ is the group velocity component in x direction. The $i$ is the phonon state. For boundary emission, a particle will be emitted from hot boundary if a random number $R_1$ is smaller than $\varepsilon_1\left(\varepsilon_1+\varepsilon_2\right)^{-1}$. Otherwise, the particle will be emitted from the cold boundary. Then, the phonon state of the particle is determined to be state i if another random number $R_2$ satisfies
\begin{align} \label{eq:S3}
\frac{\sum_{j}^{i}p_j}{\sum_{j}p_j} \le R < \frac{\sum_{j}^{i+1}p_j}{\sum_{j}^{i}p_j}
\end{align}
where $p_j=\partial e_j^{\text{eq}}/{\partial T}\left[\left(T_1-T_{\text{eq}}\right)+\left(-dT/dx\right)_1\tau_jv_{j,x}\right]v_{j,x}$ for the left boundary. The summation in Eq. (S3) is over all modes $j$ that have a positive group velocity along the $x$-direction. For the cold boundary, $p_j=\partial e_j^{\text{eq}}/{\partial T}\left[\left(T_2-T_{\text{eq}}\right)+\left(-dT/dx\right)_2\tau_jv_{j,x}\right]v_{j,x}$ and the summation is over all  $v_{j,x}<0$ modes.

The sampling particles are allowed to freely move with their group velocities between two scattering events. The time of free flight until a scattering event, $\Delta{t}$, is stochastically determined with a random number R as $\Delta{t}=-\tau_i\ln{R}$. When a three-phonon scattering event occurs, the phonon state i after the scattering is found such that another random number ($R'$) satisfies the following condition:
\begin{align} \label{eq:S4}
\frac{\sum_{j}^{i}\frac{de_j^{\text{eq}}}{dT}\frac{1}{\tau_j}}{\sum_{j}\frac{de_j^{\text{eq}}}{dT}\frac{1}{\tau_j}} \le R' < \frac{\sum_{j}^{i+1}\frac{de_j^{\text{eq}}}{dT}\frac{1}{\tau_j}}{\sum_{j}\frac{de_j^{\text{eq}}}{dT}\frac{1}{\tau_j}}
\end{align}
The amount of time that the particle spend in each control volume is recorded until the particle leave the computational domain. The phonon distribution in each control volume is then determined from the time that sample particles spent in the control volume:
\begin{align} \label{eq:S5}
e_i^{\text{d}}=\frac{\dot{\epsilon}_{\text{eff}}V_{\text{uc}}t_i}{NL}
\end{align}
where the $\dot{\epsilon}_{\text{eff}}=(\dot{\epsilon}_1+\dot{\epsilon}_2)/2$ and $t_i$ is the total time that sampling particles with phonon state $i$ spent in the control volume. The $L$ is the length of one control volume. More details of the kinetic Monte Carlo simulation can be found in a previous study[10].

\subsection{\textbf{Local deviational temperature and heat flux}}
\label{S2.1}
In our simulation, the local deviational temperature in each control volume is calculated as
\begin{align} \label{eq:S6}
T_{\text{loc}}^{\text{d}} = \left( C_{\text{V}} N V_{\text{uc}} \right) ^{-1} \sum_{i} e_{i}^{\text{d}} 
\end{align}
where $e_i^{\text{d}}$ is the deviational energy distribution from the global equilibrium, $\hbar\omega f-\hbar\omega f^{\text{eq}}$ , and $C_{\text{V}}$ is the volumetric specific heat. The heat flux $q''$ is defined as
\begin{align} \label{eq:S7}
q'' = \left( N V_{\text{uc}} \right) ^{-1} \sum_{i} v_{i,x} e_{i}^{\text{d}}
\end{align}
The $q^+$ and $q^-$, which are the heat flux from the phonon modes with positive and negative group velocities along the heat flux direction, can be expressed as
\begin{align} \label{eq:S8}
q^+ = \left( N V_{\text{uc}} \right) ^{-1} \sum_{i,v_{i,v_x>0}} v_{i,x} e_{i}^{\text{d}}
\end{align}
\begin{align} \label{eq:S9}
q^- = \left( N V_{\text{uc}} \right) ^{-1} \sum_{i,v_{i,v_x<0}} v_{i,x} e_{i}^{\text{d}}
\end{align}
The asymmetry of heat flux ($\Delta{q}_{\text{asym}}$) is defined as $\left|q^+-q^-\right|/\left(q^++q^-\right)$. With the bulk distribution, the $q_{\text{bulk}}^+$ and $q_{\text{bulk}}^-$ are the same:
\begin{align} \label{eq:S10}
q^+_{\text{bulk}} = \left( N V_{\text{uc}} \right) ^{-1} \sum_{i,v_{i,v_x>0}} e_{i}^{\text{loc}}+v^2_{i,x}\tau_i \frac{de_i^{\text{eq}}}{dT} (-\frac{dT}{dx})
\end{align}
\begin{align} \label{eq:S11}
q^-_{\text{bulk}} = \left( N V_{\text{uc}} \right) ^{-1} \sum_{i,v_{i,v_x<0}} e_{i}^{\text{loc}}+v^2_{i,x}\tau_i \frac{de_i^{\text{eq}}}{dT} (-\frac{dT}{dx})
\end{align}
Thus, the asymmetry of heat flux is zero when the distribution function is the bulk distribution.

\subsection{\textbf{Length of finite leads}}
\label{S2.2}

Our simulation replaces the semi-infinite leads with finite leads that emit phonons following the bulk distribution. The length of the finite leads should be sufficiently long for the phonons leaving the computational domain to follow the bulk distribution. We ran multiple simulations with different lengths of Si and Ge leads. The interfacial resistances ($R_{\text{int}}$) are shown with respect to the lengths of finite leads in Fig. S1(a). When the lead is longer than 200$\Lambda_{\text{avg}}$, the $R_{\text{int}}$ is converged showing that 200$\Lambda_{\text{avg}}$ is sufficiently long for the finite lead to be equivalent to a semi-infinite lead. In Fig. S1(b), we also present the thermal conductivity values that are spatially averaged over three control volumes near the boundaries ($\kappa_{\text{b}}$). The thermal conductivity value is converged to the intrinsic bulk thermal conductivity value when the lead is longer than 200$\Lambda_{\text{avg}}$. Thus, when the lead length is larger than 200$\Lambda_{\text{avg}}$, the phonon transport near the boundaries is similar to that in bulk case, indicating that finite leads with the length of 200$\Lambda_{\text{avg}}$ is sufficiently long to relax the phonon distribution to the bulk distribution and can replace the semi-infinite leads. The values of $\Lambda_{\text{avg}}$ for each case are shown in Table S1. 

%% figure S1
\begin{figure}[H]
\includegraphics[width=\textwidth]{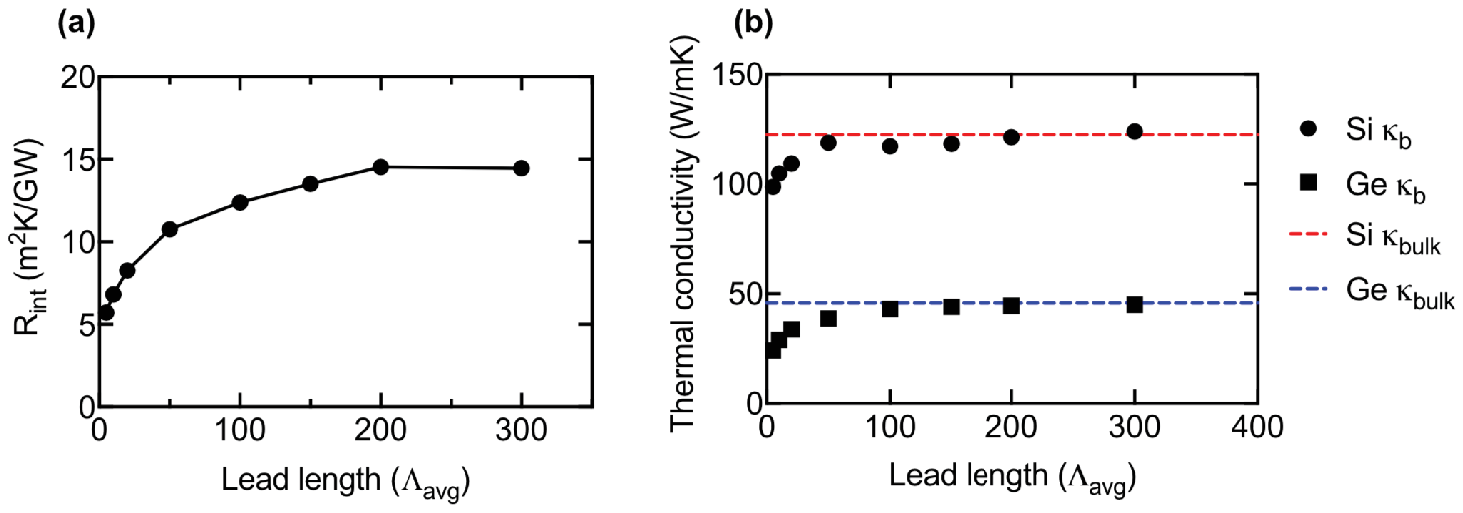}
\caption{(a) Interfacial thermal resistance in different lead lengths. (b) Spatially averaged thermal conductivity of three control volumes near the boundaries of Si (black circles) and Ge (black squares). The dashed lines are analytically calculated bulk thermal conductivity of Si (red) and Ge (blue).}
\label{fig:S1}
\end{figure}

%% Table S1
\begin{table}[H]
\centering
\begin{tabular}{||c c c||} 
 \hline
 Interface type & $\Lambda_{\text{avg}}$ of Si [nm] & $\Lambda_{\text{avg}}$ of Ge [nm] \\ [1ex] 
 \hline\hline
 Si-Ge at 100 K & 480.0 & 119.3  \\
 Si-Ge at 200 K & 58.4 & 31.3  \\
 Si-Ge at 300 K & 26.9 & 18.3  \\
 Si-Ge at 400 K & 17.6 & 13.1  \\
 Si-Ge at 500 K & 13.2 & 10.2  \\
 Si-Ge at 600 K & 10.6 & 8.4  \\
 $^{96}$Si-Ge & 13.1 & 18.3  \\
 $^{383}$Si-Ge & 6.3 & 18.3  \\
 Si-Ge' & 26.9 & 36.5  \\
 Si-Ge'' & 26.9 & 20.4  \\
 Si-Ge at 100 K ($\tau$-mod) & 26.9 & 18.3 \\
 Si$^*$-Ge$^*$ & 0.1 & 0.1  \\ [1ex]
 \hline
\end{tabular}
\caption{Mode specific heat-averaged phonon mean free path for each case}
\label{table: S1}
\end{table}

\subsection{\textbf{Entropy generation rate and its uncertainty}}
\label{S2.3}

Calculating the entropy generation rate using Eq. (11) may require extremely large number of sampling particles. One approach to obtain high fidelity entropy generation rate using relatively small number of sampling particles is to use the fact that the expression for the entropy generation rate, Eq. (11), is similar to the variance of distribution function and a variance is proportional to $N_{\text{p}}^{-1}$ where $N_p$ is the number of samples. Figure S2(a) shows the calculated entropy generation rates in a control volume in Si and Ge leads using the different number of sampling particles. Fitting with the function $N_{\text{p}}^{-1}$ shows the excellent quality with $R^2>0.99$. Thus, for each control volume, we extrapolate the entropy generation rate using $N_{\text{p}}^{-1}$ in the limit of $N_{\text{p}}^{-1}$ approaching to zero. Figures S2(b-c) show the distribution of local entropy generation with error bars in the Si and Ge leads at 300 K. The error bar based on the 95\% confidence interval of linear extrapolation is sufficiently small to capture the variation of local entropy generation rate in the leads. In particular, the error bar in Ge side is too small to be observed. 

The error bar of the local entropy generation is used to estimate the uncertainty of the decomposed resistance, $R_{\text{neq},\text{Si}}$, $R_{\text{neq},\text{Ge}}$, and $R_{\text{int}}^0$, using the uncertainty propagation method. The $R_{\text{neq},\text{Si}}$ and $R_{\text{neq},\text{Ge}}$ are calculated by integrating the local entropy generation in space as shown in Eq. (14). For the integration, the local entropy generation rate is fitted in a real space with an exponential function, $\dot{S}_{\text{gen}}\left(x\right)=\dot{S}_{\text{bulk}}+Aexp\left(-Bx\right)$. As shown in Figs. S2(b-c), the exponential function shows good fitting quality. We also calculated the $R_{\text{neq}}$ values by summing the discrete local entropy generation values without the exponential fitting. The calculated $R_{\text{neq}}$ shows only difference of 2.75\% compared to the $R_{\text{neq}}$ from the integration of the exponential function. The uncertainties of $R_{\text{neq},\text{Si}}$ and $R_{\text{neq},\text{Ge}}$ can be estimated from the 95\% confidence interval of exponential fitting parameters $A$ and $B$ with the uncertainty propagation method. The standard deviation of the fitting parameter $A$, $\sigma_A$, can be found using[11]
\begin{align} \label{eq:S12}
\sigma_A =  \sqrt{\frac{N_{\text{CV}}(u-l)}{2\text{tinv}}(0.95,N_{\text{CV}}-2)}
\end{align}
where $N_{\text{CV}}$ is the number of control volumes, which is the number of data points for exponential fitting. The $u$ and $l$ are upper limit and lower limit of confidence interval, and tinv() is the inverse of normal distribution with the 95\% confidence level and the degree-of-freedom of fitting error being $N_{\text{CV}}-2$. The $\sigma_B$ can be found using the same method. Then, the standard deviation of $R_{\text{neq}}$, $\sigma_{R_{\text{neq}}}$, can be found as[12, 13]
\begin{align} \label{eq:S13}
\sigma_{R_{\text{neq}}} =  \sqrt{(\frac{\partial R_{\text{neq}}}{\partial A}\sigma_A)^2+(\frac{\partial R_{\text{neq}}}{\partial B}\sigma_B)^2}
\end{align}
where $R_{\text{neq}}$ is the function of the fitting parameters $A$ and $B$, i.e.,
\begin{align} \label{eq:S14}
R_{\text{neq}} = (\frac{T}{q''})^2 \int_{\text{lead}} Aexp(-Bx)
\end{align}
Then, the error bar in the decomposed interfacial thermal resistance in the main manuscript, which is $\sigma_{R_{\text{int}}^0}$, can be calculated from the $\sigma_{R_{\text{neq},\text{Si}}}$ and $\sigma_{R_{\text{neq},\text{Si}}}$ as
\begin{align} \label{eq:S15}
\sigma_{R_{\text{int}}^0} = \sqrt{(\sigma_{R_{\text{neq},\text{Si}}})^2+(\sigma_{R_{\text{neq},\text{Ge}}})^2}
\end{align}
Here we ignore the uncertainty of $R_{\text{int}}$ which is the sum of $R_{\text{int}}^0$, $R_{\text{neq},\text{Si}}$, and $R_{\text{neq},\text{Ge}}$. The $R_{\text{int}}$ is calculated by simply subtracting $R_{\text{bulk}}$ from $R_{\text{tot}}$, both of which have extremely small uncertainty with the number of sample particles we used.

%% figure S2
\begin{figure}[H]
\includegraphics[width=\textwidth]{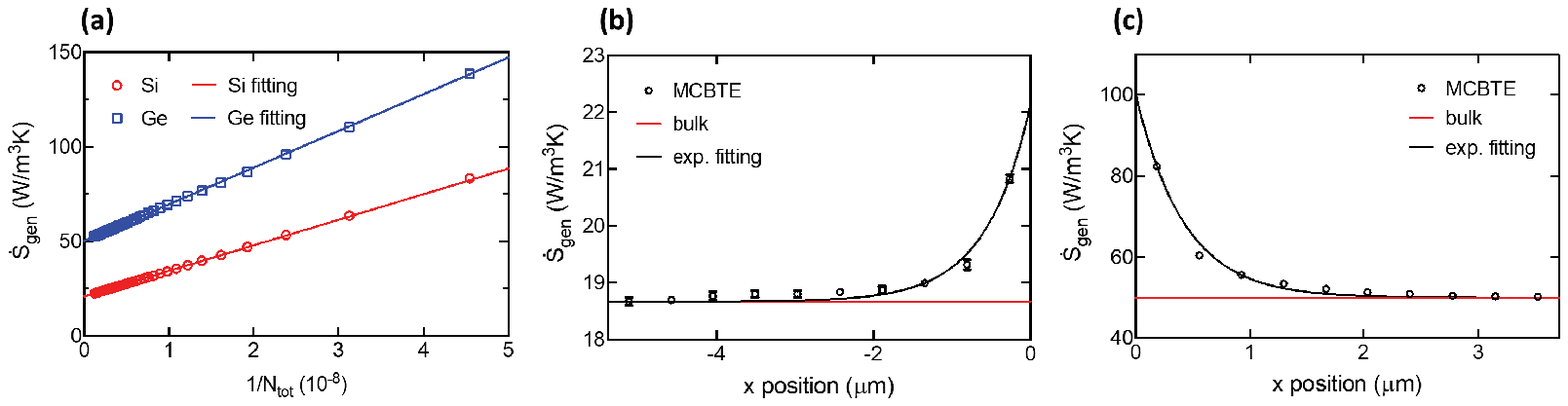}
\caption{(a) Entropy generation rates of the control volume of Si near the interface (red circles) and the control volume of Ge near the boundary (blue squares) with different number of particles. The solid line is fitting with $N_{\text{p}}^{-1}$. (b, c) Entropy generation rates in each control volume with error bar in Si and Ge at $N_{\text{p}}^{-1}=0$. The black lines in (b) and (c) are the exponential fitting of the open circles. The red lines in (b) and (c) are the rate of local entropy generation rate in infinitely large Si and Ge ($\dot{S}_{\text{bulk}}$).}
\label{fig:S2}
\end{figure}

\subsection{\textbf{Relation between local resistivity and local entropy generation rate}}
\label{S2.4}

Here we confirm the relation between $R'_{\text{neq}}$ and $\dot{S}$, $R'_{\text{neq}}=(T/q'')^2\dot{S}$, using the case of thermal transport in bulk material where the analytic solution of the PBE is known. The phonon distribution is
\begin{align} \label{eq:S16}
f_i^{\text{bulk}} = f_i^0+v_i\tau_i\frac{\partial f_i^0}{\partial T} (- \frac{dT}{dx})
\end{align}
The corresponding heat flux and resistivity are
\begin{align} \label{eq:S17}
q'' = \frac{1}{N V_{\text{uc}}} \sum_{i} \hbar\omega_i v_{i} \left[ v_i\tau_i\frac{\partial f_i^0}{\partial T} (- \frac{dT}{dx}) \right]
\end{align}
\begin{align} \label{eq:S18}
R'_{\text{bulk}} = (- \frac{dT}{dx})(q'')^{-1} = (\frac{1}{N V_{\text{uc}}} \sum_{i} \hbar\omega_i v^2_{i}  \tau_i\frac{\partial f_i^0}{\partial T})^{-1}
\end{align}
Using Eq. (11) and Eq. (S16), the rate of entropy generation is
\begin{align} \label{eq:S19}
\dot{S}=\frac{k_{\text{B}}}{N V_{\text{uc}}} \sum_{i} \frac{\left[ v_i\tau_i\frac{\partial f_i^0}{\partial T} (- \frac{dT}{dx}) \right]^2}{f_i^0(f_i^0+1)\tau_i}
\end{align}
The relation $R'_{\text{neq}}=(T/q'')^2\dot{S}$ can be confirmed using the expression for $q"$, $R'_{\text{bulk}}$, and $\dot{S}$ in Eqs. (S17-S19) with $f_i^0(f_i^0+1)=(k_{\text{B}}T^2/\hbar\omega_i)(\partial f_i^0/\partial T)$.

\section{Fictitious Si-Ge interfaces}
\label{S3}
In Fig. S3, we present the phonon dispersion of Si-Ge, $^{96}$Si-Ge and $^{383}$Si-Ge, and the phonon density-of-states (DOS) of fictitious Si. In Fig. S4, we plot the scattering rates of original Si and Ge at 300K. In Fig. S5, we show the phonon transmissivity of fictitious Si-Ge interfaces from DMM. 

%% figure S3
\begin{figure}[H]
\includegraphics[width=\textwidth]{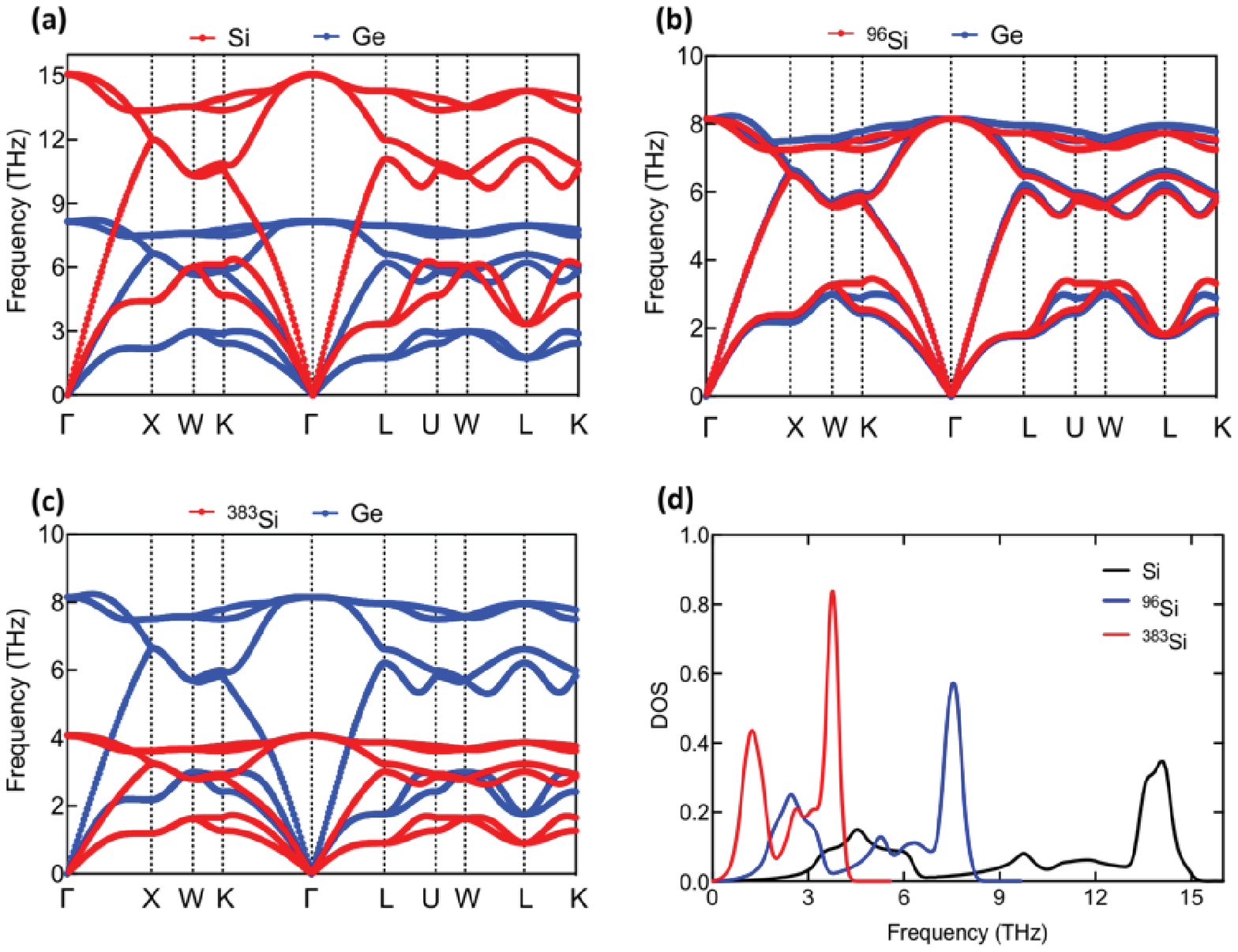}
\caption{Phonon dispersion of Si-Ge, $^{96}$Si-Ge and $^{383}$Si-Ge and phonon density-of-states (DOS) of different fictitious Si. (a-c) phonon dispersion of original Si-Ge, $^{96}$Si-Ge, and phonon dispersion of $^{383}$Si-Ge. The red and blue lines are for Si and Ge, respectively. (d) normalized phonon DOS of Si (black), $^{96}$Si (blue) and $^{383}$Si (red). }
\label{fig:S3}
\end{figure}

%% figure S4
\begin{figure}[H]
\includegraphics[width=\textwidth]{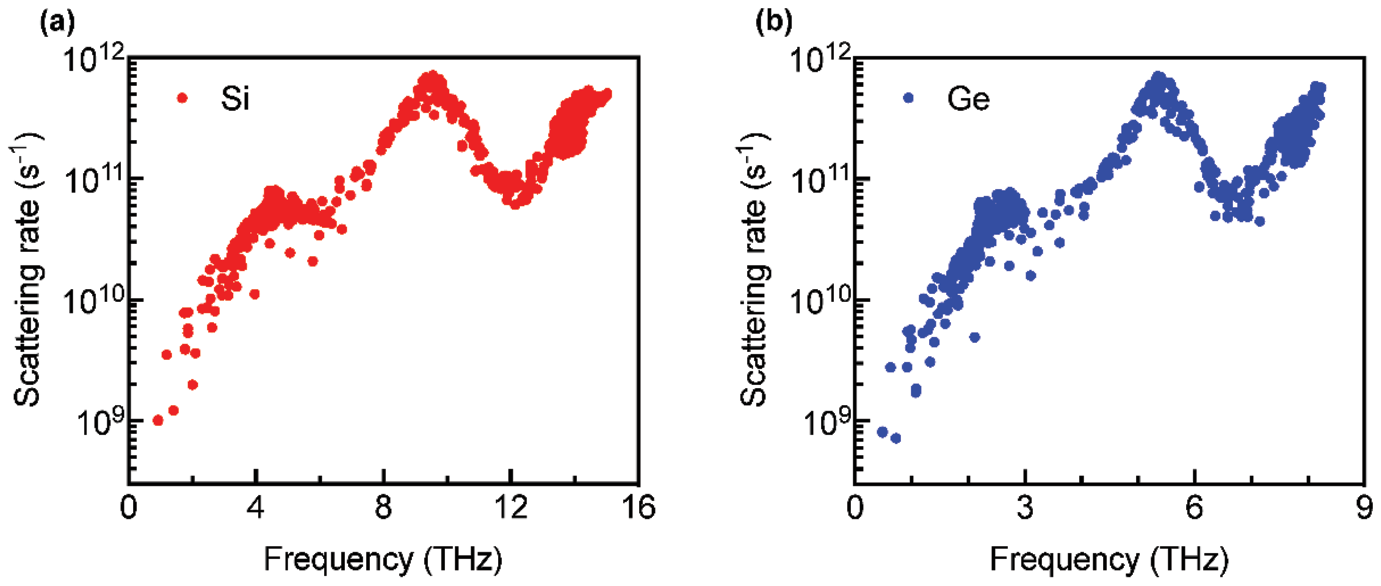}
\caption{Three-phonon scattering rates at 300 K for (a) Si and (b) Ge.}
\label{fig:S4}
\end{figure}

%% figure S5
\begin{figure}[H]
\includegraphics[width=\textwidth]{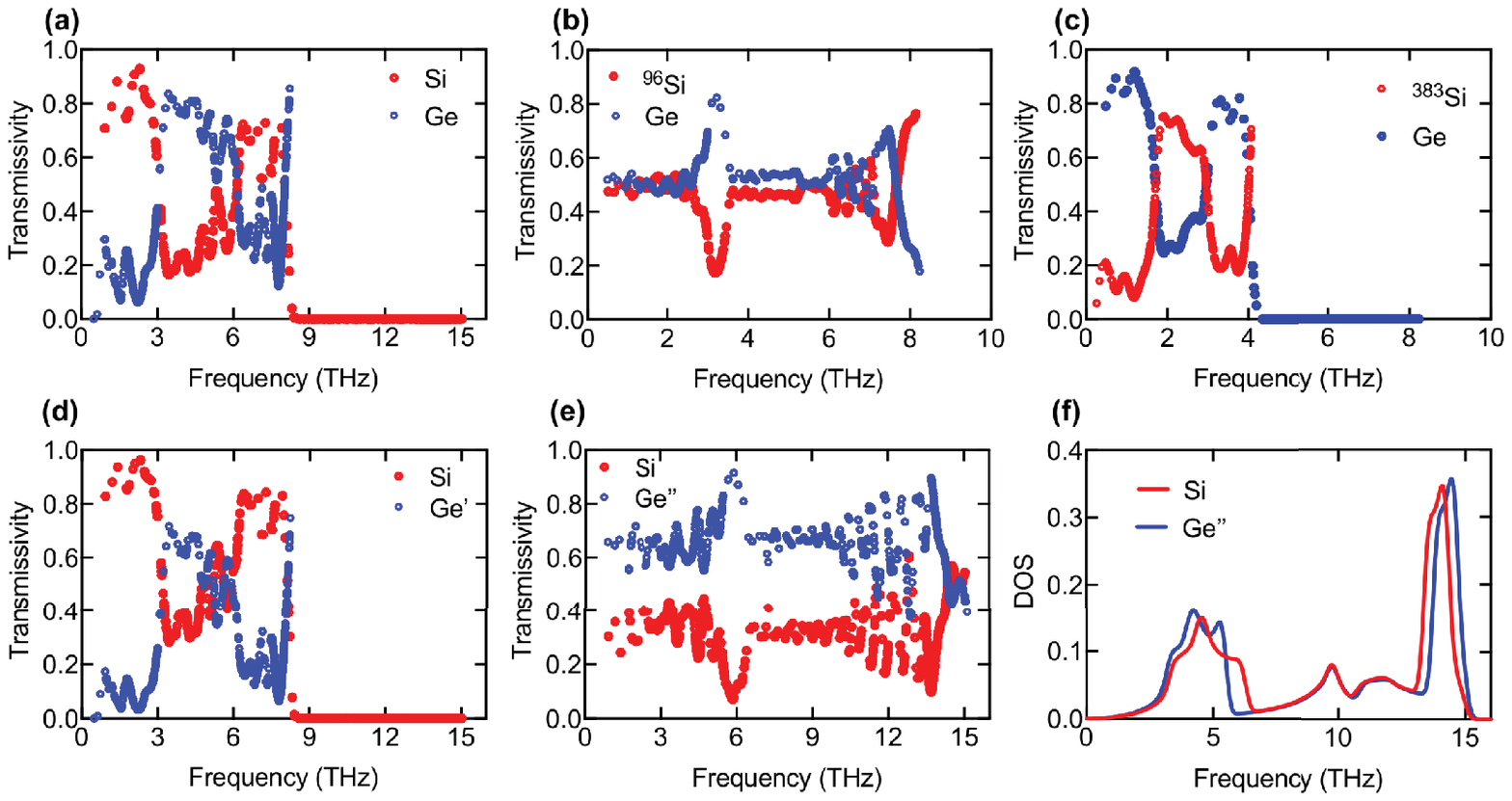}
\caption{Phonon transmissivity from DMM for different Si-Ge interfaces and phonon DOS of Si and Ge''. (a-e) Phonon transmissivity in original Si-Ge, $^{96}$Si-Ge, $^{383}$Si-Ge, Si-Ge', and Si-Ge'' interfaces. The red represents transmissivity from Si to Ge, and the blue represents transmissivity from Ge to Si. (f) Comparable phonon DOS of Si (red) and Ge'' (blue).}
\label{fig:S5}
\end{figure}

\newpage
\section*{References}
\label{ref}

[1] W. Li, J. Carrete, N. A. Katcho, N. Mingo, ShengBTE: A solver of the Boltzmann transport equation for phonons, Computer Physics Communications, 185 (2014) 1747-1758.

[2] G. Kresse, J. Hafner, Ab initio molecular dynamics for liquid metals, Phys Rev B Condens Matter, 47 (1993) 558-561.

[3] G. Kresse, J. Hafner, Ab initio molecular-dynamics simulation of the liquid-metal-amorphous-semiconductor transition in germanium, Phys Rev B Condens Matter, 49 (1994) 14251-14269.

[4] G. Kresse, J. Furthmüller, Efficiency of ab-initio total energy calculations for metals and semiconductors using a plane-wave basis set, Computational materials science, 6 (1996) 15-50.

[5] G. Kresse, J. Furthmüller, Efficient iterative schemes for ab initio total-energy calculations using a plane-wave basis set, Physical review B, 54 (1996) 11169.

[6] G. Kresse, J. Furthmüller, J. Hafner, Theory of the crystal structures of selenium and tellurium: The effect of generalized-gradient corrections to the local-density approximation, Phys Rev B Condens Matter, 50 (1994) 13181-13185.

[7] G. Kresse, D. Joubert, From ultrasoft pseudopotentials to the projector augmented-wave method, Physical review b, 59 (1999) 1758.

[8] A. Togo, I. Tanaka, First principles phonon calculations in materials science, Scripta Materialia, 108 (2015) 1-5.

[9] J.-P.M. Péraud, N.G. Hadjiconstantinou, Efficient simulation of multidimensional phonon transport using energy-based variance-reduced Monte Carlo formulations, Physical Review B, 84 (2011).

[10] J.-P.M. Péraud, N.G. Hadjiconstantinou, An alternative approach to efficient simulation of micro/nanoscale phonon transport, Applied Physics Letters, 101 (2012).

[11] J.P. Higgins, T. Li, J.J. Deeks, Choosing effect measures and computing estimates of effect, Cochrane handbook for systematic reviews of interventions, DOI (2019) 143-176.

[12] L.A. Goodman, On the exact variance of products, Journal of the American statistical association, 55 (1960) 708-713.

[13] H.H. Ku, Notes on the use of propagation of error formulas, Journal of Research of the National Bureau of Standards, 70 (1966) 263-273.